\documentclass{tufte-handout}
\usepackage[utf8]{inputenc}
\usepackage{graphicx}
\usepackage{stackrel,amsmath,amssymb,amsfonts,amsthm,bm}
\usepackage{nicematrix}
\usepackage{mathtools}
\usepackage{cancel}
\newcommand\Ccancel[2][black]{\renewcommand\CancelColor{\color{#1}}\cancel{#2}}

\usepackage{tikz-cd}
\usepackage{tikz}
\usepackage{tkz-euclide}
\usepackage{pgfplots}
\usepackage{tikz-3dplot}
\usetikzlibrary{shapes}
\usetikzlibrary{shapes.geometric}
\usetikzlibrary{calc}
\usetikzlibrary{fit}
\usetikzlibrary{positioning}
\usetikzlibrary{decorations.markings,arrows,arrows.meta,bending}
\usetikzlibrary{decorations.pathreplacing}
\usetikzlibrary{decorations.pathmorphing}
\usetikzlibrary{decorations.markings}
\usetikzlibrary{arrows.meta,bending}
\usetikzlibrary{math} 
\tikzset{->-/.style={decoration={markings,
  mark=at position #1 with {\arrow[line width=2pt]{>}}},postaction={decorate}}}

\DeclareMathOperator{\im}{im}

\DeclareMathOperator{\Ima}{Im}
\DeclareMathOperator{\grad}{grad}
\DeclareMathOperator{\Div}{div}
\DeclareMathOperator{\curl}{curl}

\tikzset{
  pics/torus/.style n args={3}{
    code = {
      \providecolor{pgffillcolor}{rgb}{1,1,1}
      \begin{scope}[
          yscale=cos(#3),
          outer torus/.style = {draw,line width/.expanded={\the\dimexpr2\pgflinewidth+#2*2},line join=round},
          inner torus/.style = {draw=pgffillcolor,line width={#2*2}}
        ]
        \draw[outer torus] circle(#1);\draw[inner torus] circle(#1);
        \draw[outer torus] (180:#1) arc (180:360:#1);\draw[inner torus,line cap=round] (180:#1) arc (180:360:#1);
      \end{scope}
    }
  }
}

\newtheorem*{theorem}{Theorem}
\usepackage{framed}

\colorlet{shadecolor}{gray!15}

\newcommand{\leftrarrows}{\mathrel{\raise.75ex\hbox{\oalign{%
  $\scriptstyle\leftarrow$\cr
  \vrule width0pt height.5ex$\hfil\scriptstyle\relbar$\cr}}}}
\newcommand{\lrightarrows}{\mathrel{\raise.75ex\hbox{\oalign{%
  $\scriptstyle\relbar$\hfil\cr
  $\scriptstyle\vrule width0pt height.5ex\smash\rightarrow$\cr}}}}
\newcommand{\Rrelbar}{\mathrel{\raise.75ex\hbox{\oalign{%
  $\scriptstyle\relbar$\cr
  \vrule width0pt height.5ex$\scriptstyle\relbar$}}}}
\newcommand{\longleftrightarrows}{\leftrarrows\joinrel\Rrelbar\joinrel\lrightarrows}

\newenvironment{thm}
  {\begin{shaded}\begin{theorem}}
  {\end{theorem}\end{shaded}}

\theoremstyle{definition}
\newtheorem*{remark}{Remark}

\title{Combinatorial Hodge Theory in Simplicial Signal Processing - DAFx2023 Lecture Notes}
\author{Georg Essl --- University of Wisconsin - Milwaukee}
\date{November 2023}

\begin{document}

\maketitle

\section{Introduction}
These are lecture notes accompanying a tutorial presented at
DAFx2023 in Copenhagen on the topic of topology in digital signal
processing and sound synthesis and specifically on combinatorial Hodge theory in graph and simplicial signal processing. It is in a sense a continuation of material presented at DAFx2022 in Vienna, and lecture notes for this previous presentation are available already \cite{essl2022topology}. Some material on simplicial complexes, homology, and sheaves are shared between sets of lecture notes, though much of the bulk of the material covers different topics. The material repeated here is included to keep this set of lecture notes reasonably self-contained.

Topology has become an increasingly developed topic in digital signal processing as well as applied and computational topology, finite element methods, and other areas. This tutorial and its notes seek to provide a concrete, accessible introduction to topics of interest. The core of these notes is what is known as combinatorial Hodge theory. All of the mathematics used is essentially simply linear algebra, though we will occasionally take a just slightly wider scope and see a group or two, but group theory is not necessary to understand the material.

\section{Combinatorial Construction of Topological Spaces}

\begin{marginfigure}
    \centering
    \includegraphics[width=.95\marginparwidth]{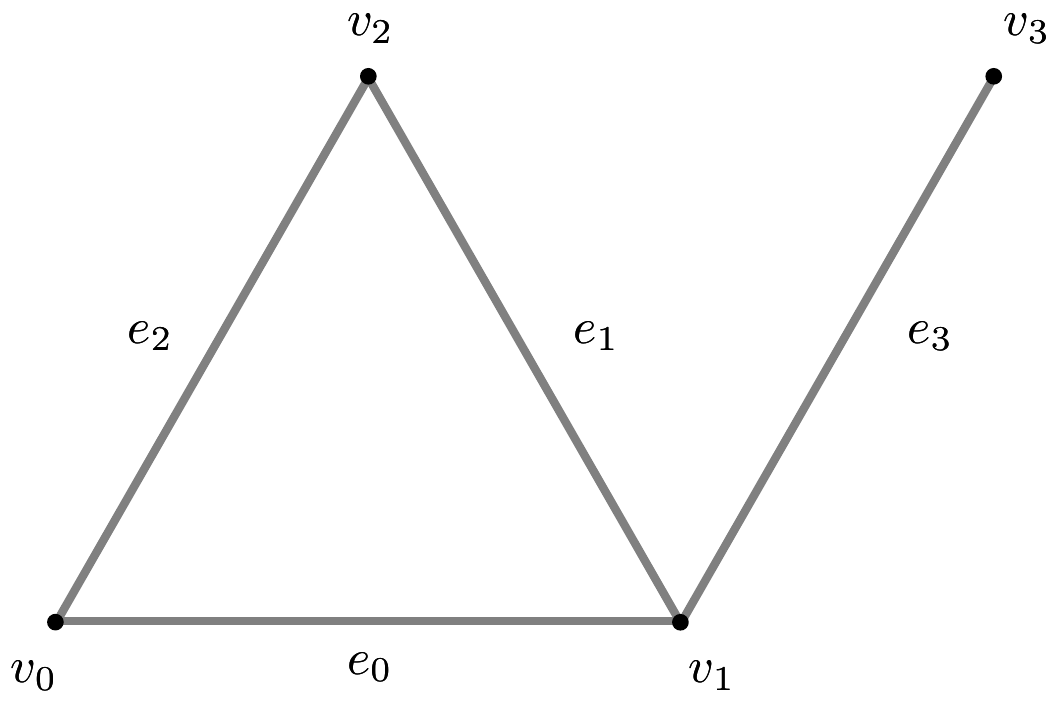}
    \caption{A simple Graph.}
    \label{fig:graph}
\end{marginfigure}

In many applications, topological spaces naturally arise as one creates interconnectivity or combinations of entities. Of models of connectivity, graphs are most familiar to practitioners in signal processing. It is customary to define graphs by their collection of vertices and edges written as $G=(V,E)$ with $V=\{v_0,v_1,\ldots,v_n\}$ and $E=\{e_0,e_1,\ldots,e_m\}$. An interesting observation about this definition is that it treats edges and vertices as completely separate, though one usually thinks of edges as connecting vertices. The relationship between edges and vertices is introduced via some connectivity construction. Matrices can be used to capture this information. Two commonly used examples of matrices capturing connectivity information are the adjacency and the incidence matrices. Consider the following simple graph depicted in Figure \ref{fig:graph}.
The {\em adjacency matrix} captures a vertex to vertex relationship and has $1$ as entry if an edge is present between a pair of vertices. Hence we can fill out the adjacency matrix as follows:%
							\[A_G=\bordermatrix{
  ~ & v_0 & v_1 & v_2 & v_3 \cr
  v_0 & 0 & 1 &  1 & 0 \cr
  v_1 & 1 & 0 & 1 & 1 \cr
  v_2 & 1 & 1 & 0 & 0 \cr
  v_3 & 0 & 1 & 0 & 0 \cr
}\]
The {\em incidence matrix} relates vertices to edges, and has a $1$ entry if an edge is "incident" to an edge (or vice versa!) and for the example graph we get the following matrix:%
							\[B_G=\bordermatrix{
  ~ & e_0 & e_1 & e_2 & e_3 \cr
  v_0 & 1 & 0 & 1 & 0 \cr
  v_1 & 1 & 1 & 0 & 1 \cr
  v_2 & 0 & 1 & 1 & 0 \cr
  v_3 & 0 & 0 & 0 & 1 \cr
}\]
Matrix representation immediately invites a kind of algebraization of the graphs as we can now study properties of the matrix. This is in fact a way to enter into algebraic graph theory. In fact, the incidence matrix plays a much more important role in this algebraization. We will develop some of the reasons for this later.

\subsection{Simplicial Complex as Generalization of Graphs}
\begin{figure}[h]
    \centering
    \includegraphics[width=.95\textwidth]{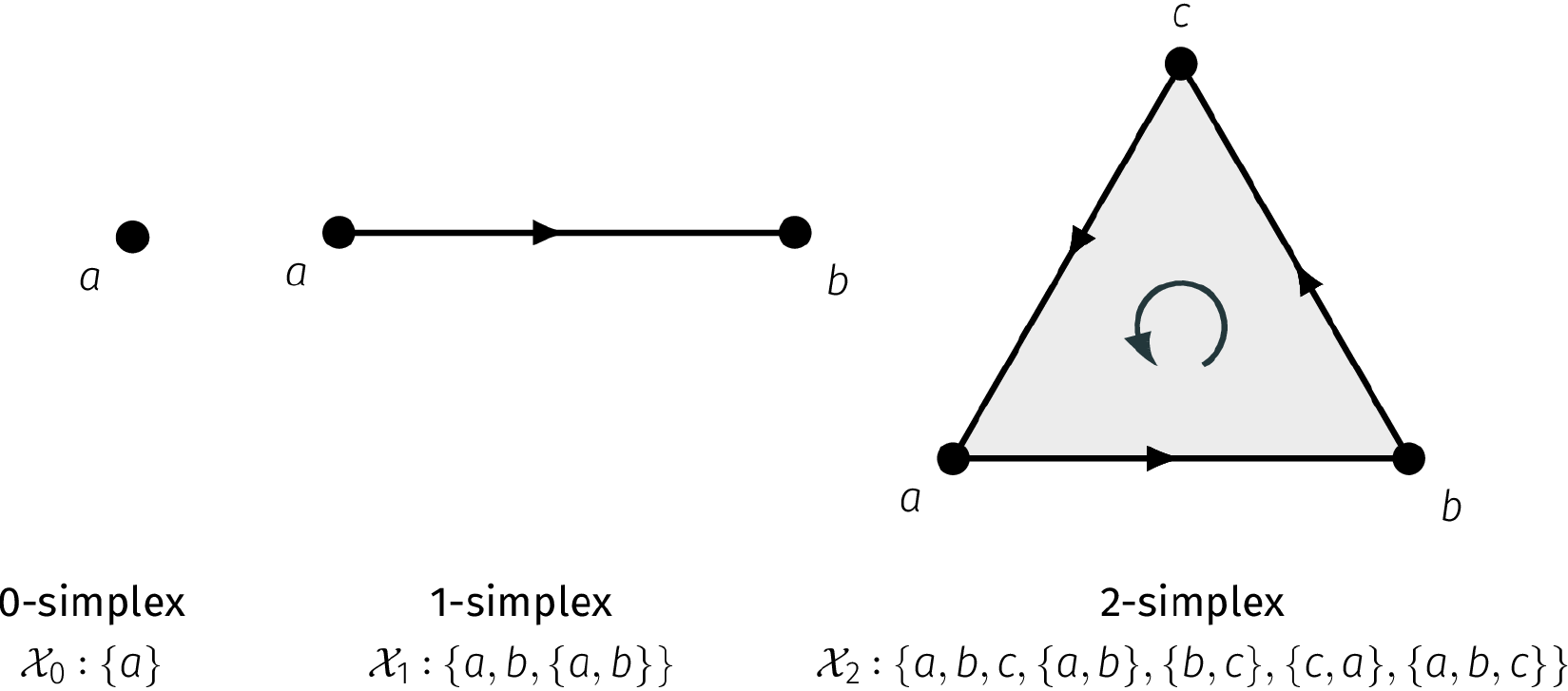}
    \caption{Low-dimensional simplices.}
    \label{fig:simplices}
\end{figure}
Simplices are higher-dimensional generalizations of vertices and lines. The next entity we need is something capturing the notion of area. The simplest possible combinatorical structure is that of an area captured by three points and surrounded by three edges (see Figure \ref{fig:simplices}).

\begin{marginfigure}[-17\baselineskip]
    \centering
    \includegraphics[width=.85\marginparwidth]{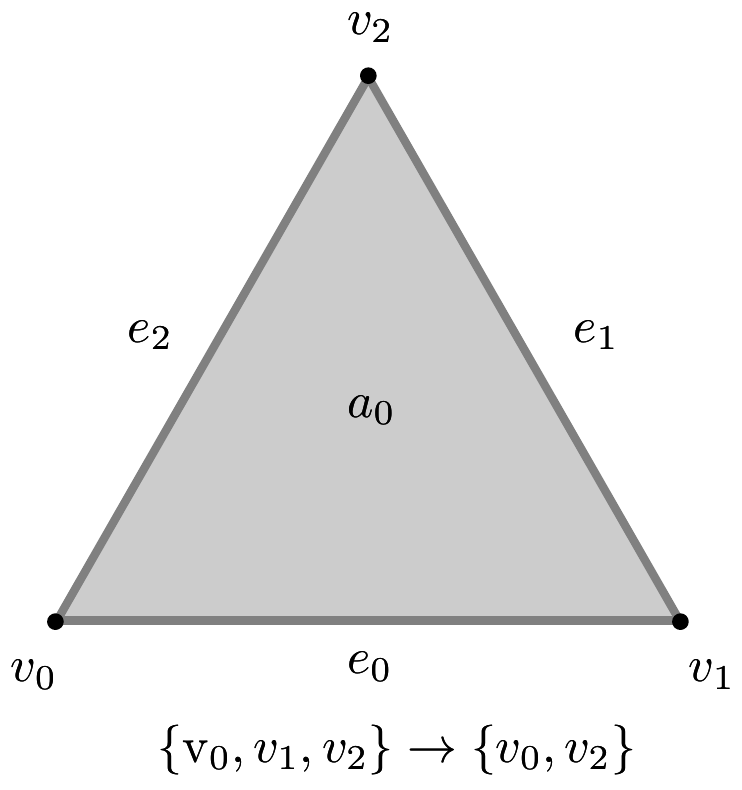}
    \includegraphics[width=.85\marginparwidth]{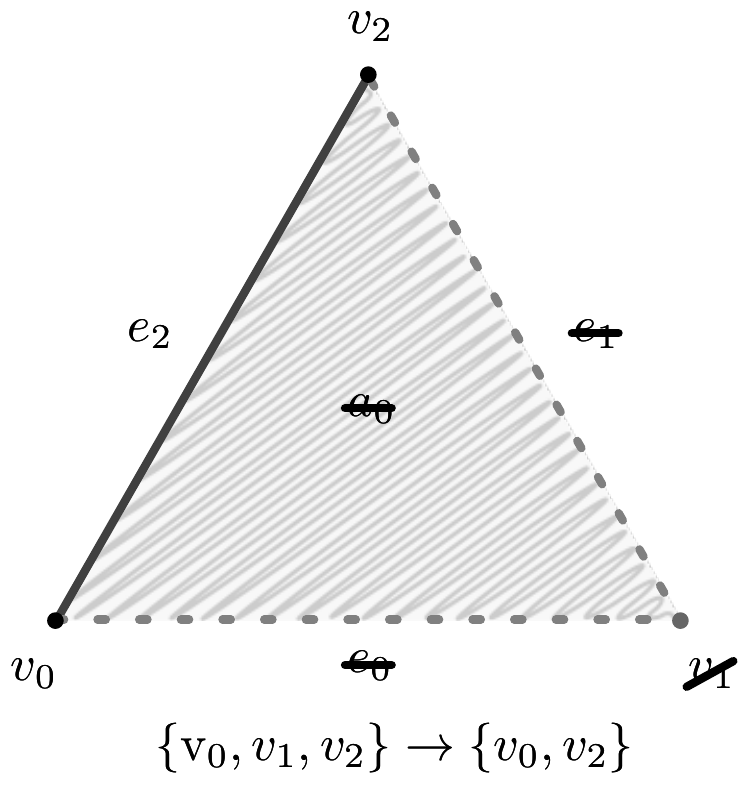}
    \caption{Deleting an entry in the set is the same as returning the face opposite the removed $0$-simplex.}
    \label{fig:facemap}
\end{marginfigure}

\begin{marginfigure}
    \centering
    \includegraphics[width=.85\marginparwidth]{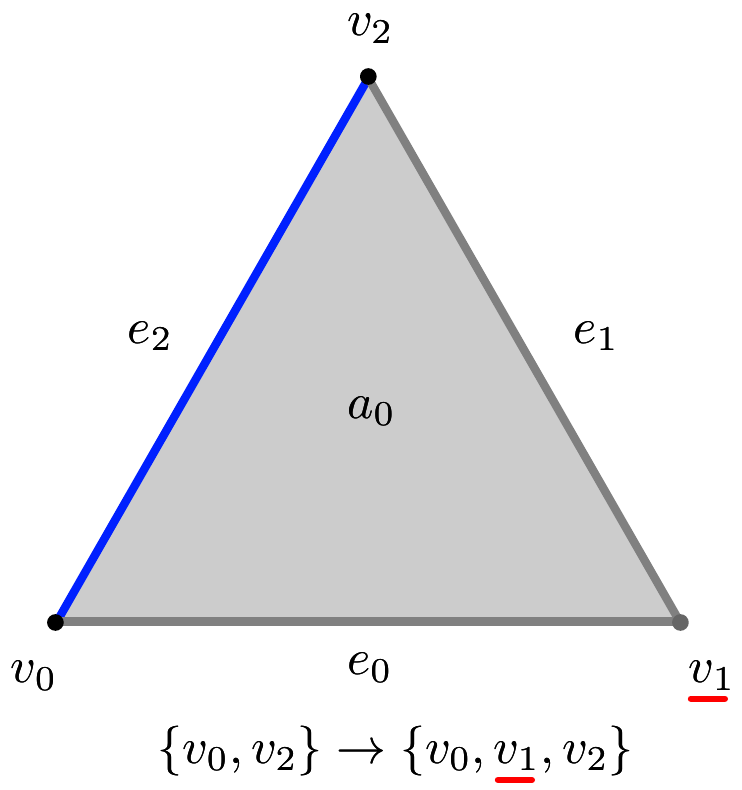}
    \caption{Inserting an entry in the set is the same as returning the coface $2$-simplex.}
    \label{fig:cofacemap}
\end{marginfigure}

This process can be continued. A volume can be constructed by surrounding it by four triangles and so forth into higher dimensions. This can be described by sets just as we did for graphs, though we will organize our sets slightly differently. This justifies giving things a new name. A vertex will be called a $0$-simplex and its set is simply a label for each vertex $\mathcal{X}_0=\{a\}$. For all higher-order simplices we require that they include the sub-simplices from which they are built. In this setup, an edge also contains the vertices it connects and is now called a $1$-simplex and its set description hence is $\mathcal{X}_1=\{a,b,\{a,b\}\}$. This is simply a somewhat different version of collecting vertex and edge information compared to our graph sets. However, we are now ready to write down the 2-simplex $\mathcal{X}_2=\{a,b,c,\{a,b\},\{b,c\},\{c,a\},\{a,b,c\}\}$. Observe that all the lower-order simplices in this set are just subsets of the largest one. This leads to data compression, because we can always construct the lower order simplices by set deletions. Set deletion also gives us a way to get sub-simplices as depicted in Figure \ref{fig:facemap}. We can get a map to a face of an $n$-simplex by deleting one entry from its set, and it will return the $n-1$-simplex opposite to the one we deleted. 
This map is hence named face-map and it takes us from an $n$-simplex to one of its $n-1$-simplex faces. Sometimes, it is convenient to go in the other direction and the map is then called coface-map \ref{fig:cofacemap} (the co-prefix refers to inverting the direction from which a map is taken). (Co)face maps are all we need to navigate a single simplex.

\begin{marginfigure}
    \centering
    \includegraphics[width=.95\marginparwidth]{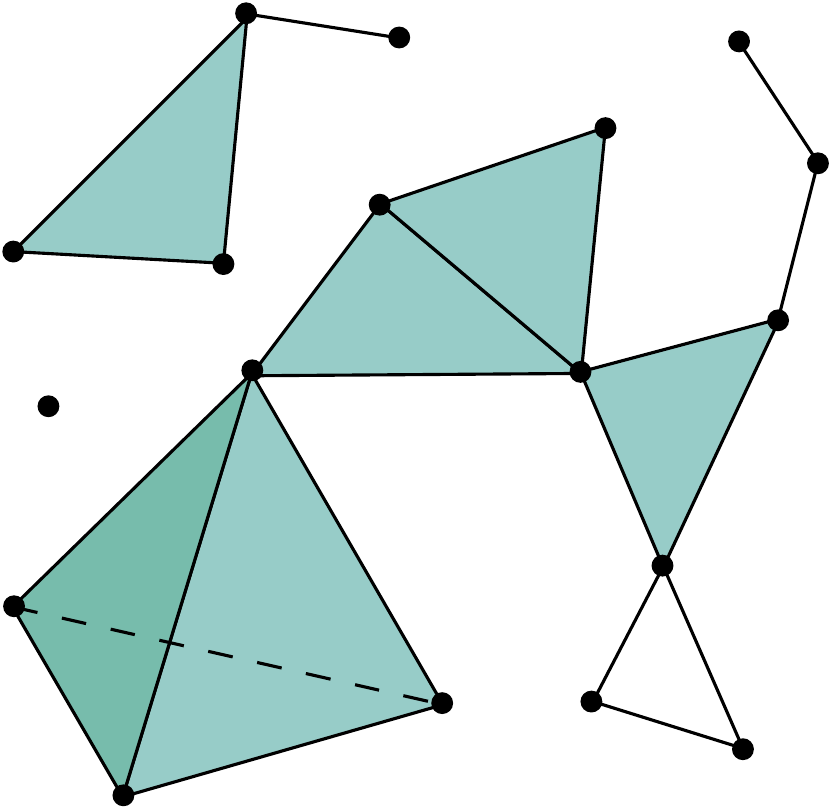}
    \caption{A simplicial complex.}
    \label{fig:simplicialcomplex}
\end{marginfigure}

An abstract simplicial complex consists of simplices that are connected together by the rule that they can only connect along shared (sub)simplices as seen in Figure \ref{fig:simplicialcomplex}. The word abstract here denotes that we do not think of a particular geometric configuration of the simplices in some space, so the notion of a simplex touching or penetrating another simplex is not well defined. This should be familiar from graph theory as used for data-flow. Strictly spreaking, we are only using that two points in a flow are connected, and some graphical depiction of the connection (whether it is for example a straight, or squiggly line or arrow) is an arbitrary choice to make this connection more visual or geometric.

\section{Homology}

\begin{figure}[h]
    \centering
    \includegraphics[width=.95\textwidth]{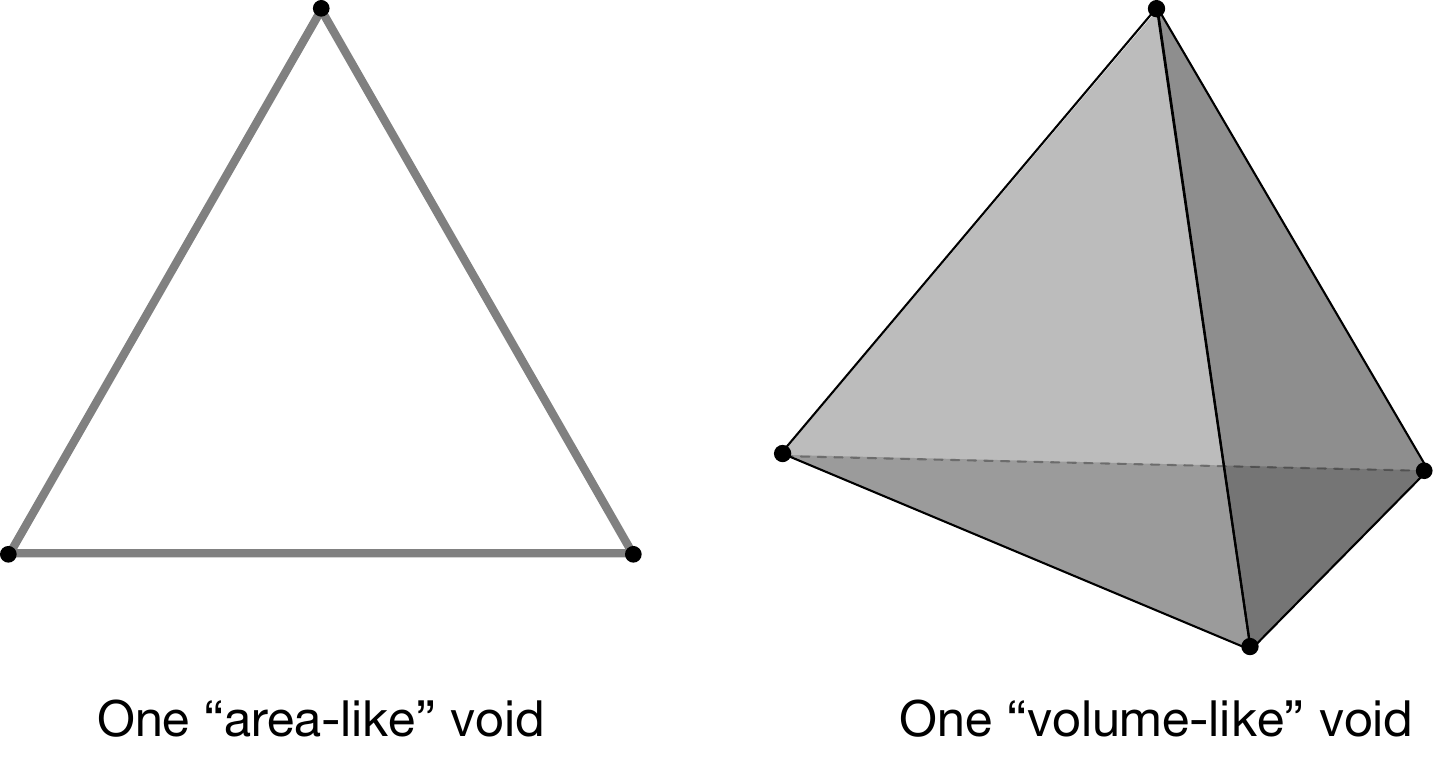}
    \caption{A triangle can either be filled or empty inside. If it is empty inside we call this a {\em void}. The inside of an empty tetrahedron is also a void but of different dimensions. The first is "area-like" the second "volume-like".}
    \label{fig:voids}
\end{figure}

\begin{figure}[h]
    \centering
    \includegraphics[width=.95\textwidth]{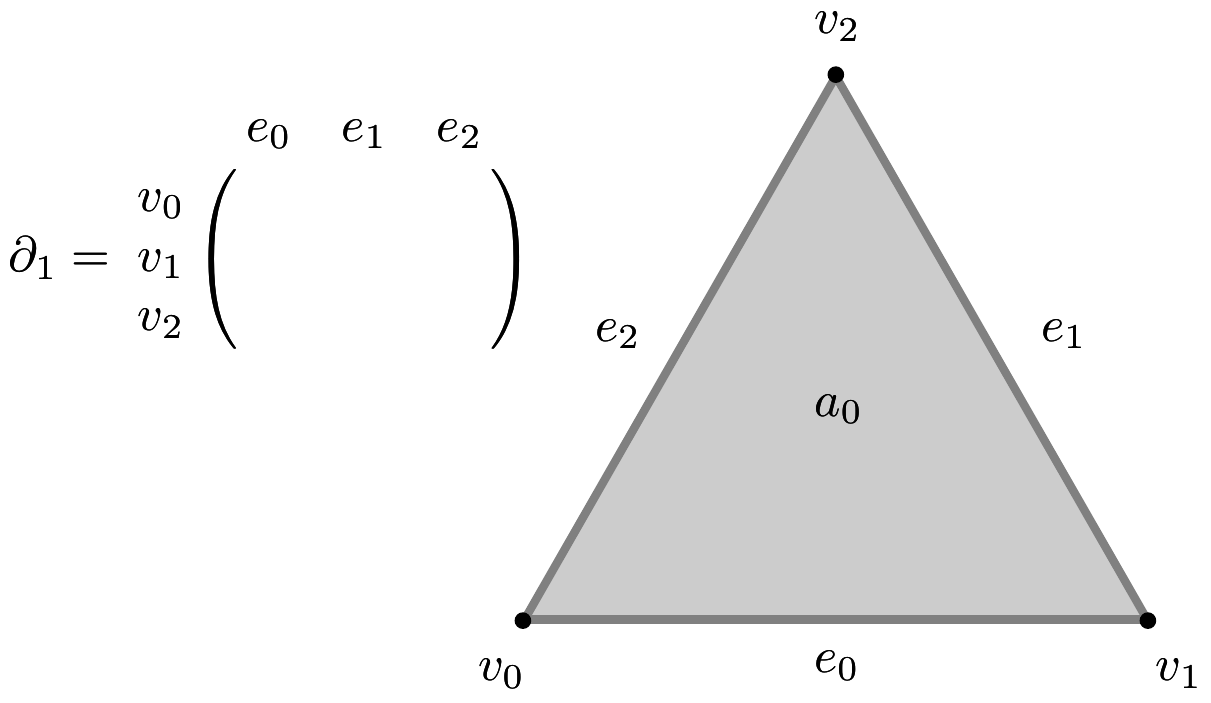}
    \caption{The boundary matrix of $1$-simplices in a simple simplicial complex relates edges to bounding vertices.}
    \label{fig:homologyempty}
\end{figure}

Homology counts $n$-dimensional voids as well as the number of connected components of a topological space. For us (and in computational settings) these topological spaces will be simplicial complexes. To understand what we mean by an $n$-dimensional void, consider the example depicted in Figure \ref{fig:voids} showing two low-dimensional examples.
A triangle bounds an empty area inside of it is not filled. This is an example of an "area-like" void, referring to the dimensionality of the simplex that would fill it. It takes three lines to fence in this void, and we observe that in fact one only gets a $n$-void if it is fully bounded by such an $n-1$ fence. We see that this also holds for the example of the tetrahedron shape (from the boundary of a $3$-simplex). The four $2$-simplices that bound it enclose a "volume-like" void inside (if it is not filled in).
It turns out that the algorithmic computations of these dimensional voids for simplicial complexes are fairly easy. Ultimately, it turns out that we compute the rank information of two matrices, called boundary matrices, to compute the number of voids. Stated like this, this may appear mysterious, but we have already noted that we only get a void when it is fully fenced in or bounded. But there is a second criterion: the fenced-in area needs to indeed be empty and not be filled with an $n$-simplex. The aim here is to construct an algebraic setup in which we capture closed boundaries, which we will call {\em cycles} and denote by the letter $Z$ (likely from the German "Zyklus"). Let's see how we get a cycle from the construction of a {\em boundary matrix}. Here we construct the boundary matrix from $1-simplices$ to $0-simplices$. Recall that each $1$-simplex in isolation is bounded by two $0$-simplices. The boundary matrix captures this information for all $1$-simplices in the simplicial complex. For simplicity, we will consider here only a single 2-simplex and its boundary matrix $\partial_1$ in Figure \ref{fig:homologyempty}.

\begin{figure}[h]
    \centering
    \includegraphics[width=.95\textwidth]{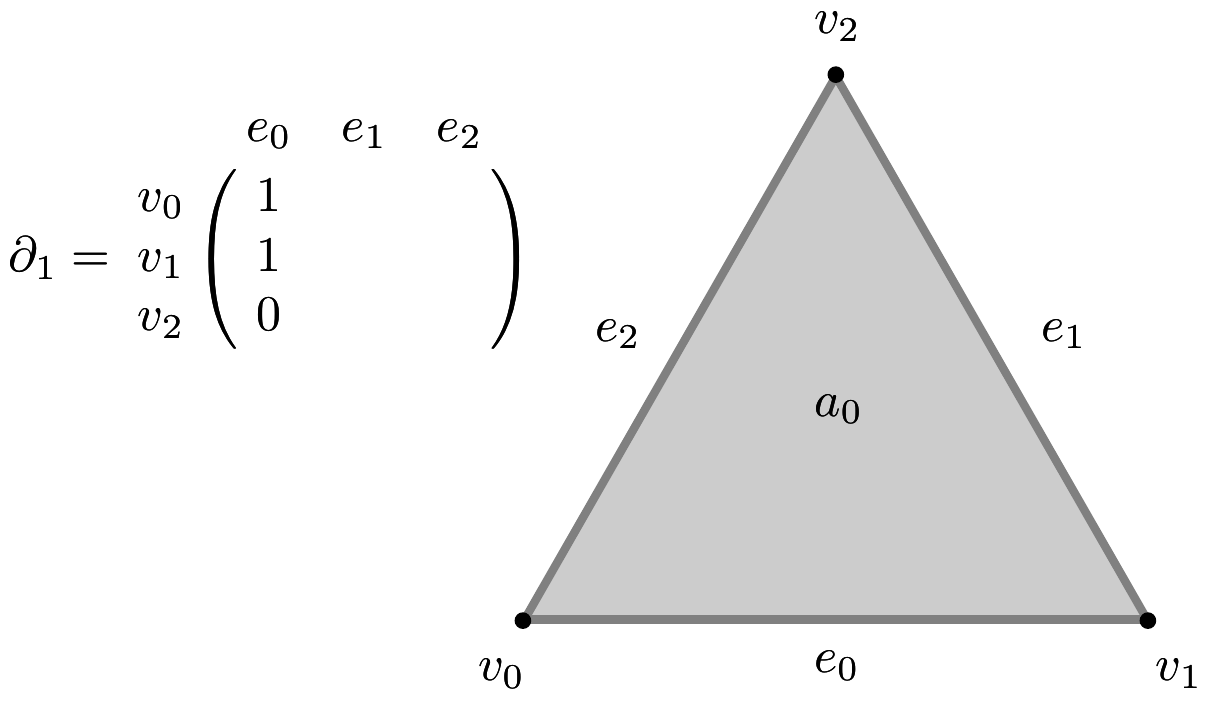}
    \caption{Two vertices $v_0$ and $v_1$ are bounding the edge $e_0$.}
    \label{fig:homologyonecolumn}
\end{figure}

\begin{figure}[h]
    \centering
    \includegraphics[width=.95\textwidth]{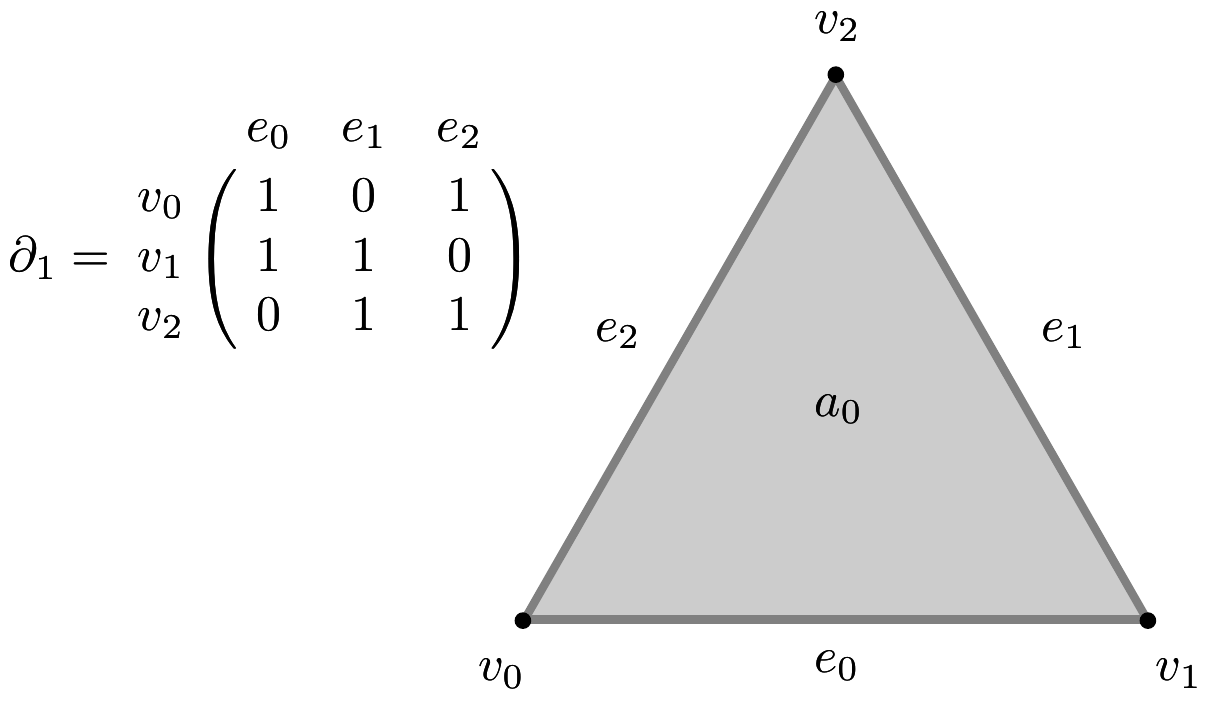}
    \caption{The completed boundary matrix from edges to vertices.}
    \label{fig:homologyfull}
\end{figure}

To populate our boundary matrix, we simply put a $1$ for each vertex $v_n$ that bounds a given edge $e_m$. We see that edge $e_0$ is bounded by vertices $v_0$ and $v_1$ and hence we get the following column entries (Figure \ref{fig:homologyonecolumn}).
We continue this process for all edges and arrive at a fully filled boundary matrix as depicted in Figure \ref{fig:homologyfull}.
While we simply entered $1$ entries here, we can make all my arguments in this simple binary setting without overflow. This means that all our additions and subtractions are carried out by \texttt{xor} operations. To mathematicians, this is known as the cyclic finite group $\mathbb{Z}/\mathbb{Z}_2$ and we can do linear algebra in this setting.
\begin{figure}[h]
    \centering
    \includegraphics[width=.95\textwidth]{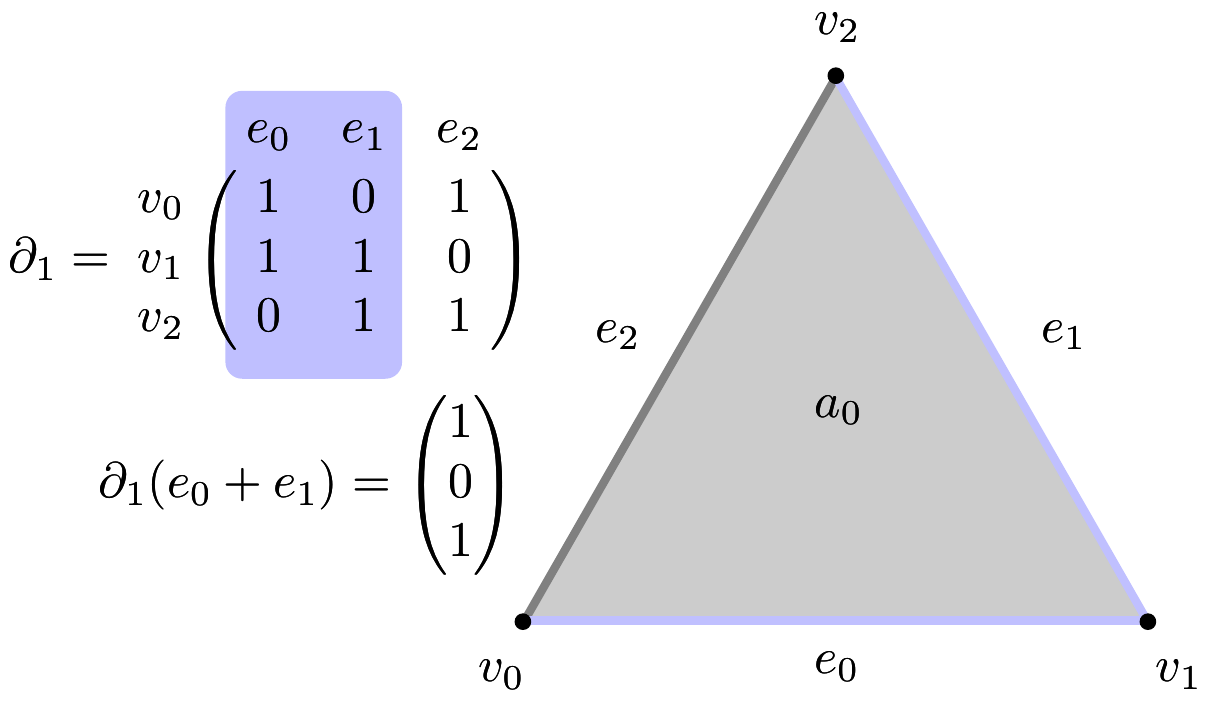}
    \caption{The linear combination of the first two columns.}
    \label{fig:homology}
\end{figure}
Given that we can do linear algebra, we can add and subtract columns (or rows). Let us add the first two columns (Figure \ref{fig:homology}). Observe that the resulting vector is actually identical to the remaining edge column $e_2$! That is, these three columns are not linearly independent. Any pair of them is, but adding in the final column is not. We will not show a figure of this, but you can easily check that the same holds for any chain of edges that eventually close on themselves in a cycle. In other words, if we sum over a closed chain of edges we get a zero vector and in our example we have $\partial_1(e_0+e_1+e_2)=(0,0,0)^T$. From linear algebra we know that linear dependence reduces rank. For every independent cycle (in the sense of linear algebra) we hence get a rank reduction. The size of the rank reduction is the size of the null-space of a linear map (here our matrix). This is the first key observation to computing homology.\marginnote{Independent cycles lead to rank reduction in the boundary matrix.}

\begin{figure}[h]
    \centering
    \includegraphics[width=.95\textwidth]{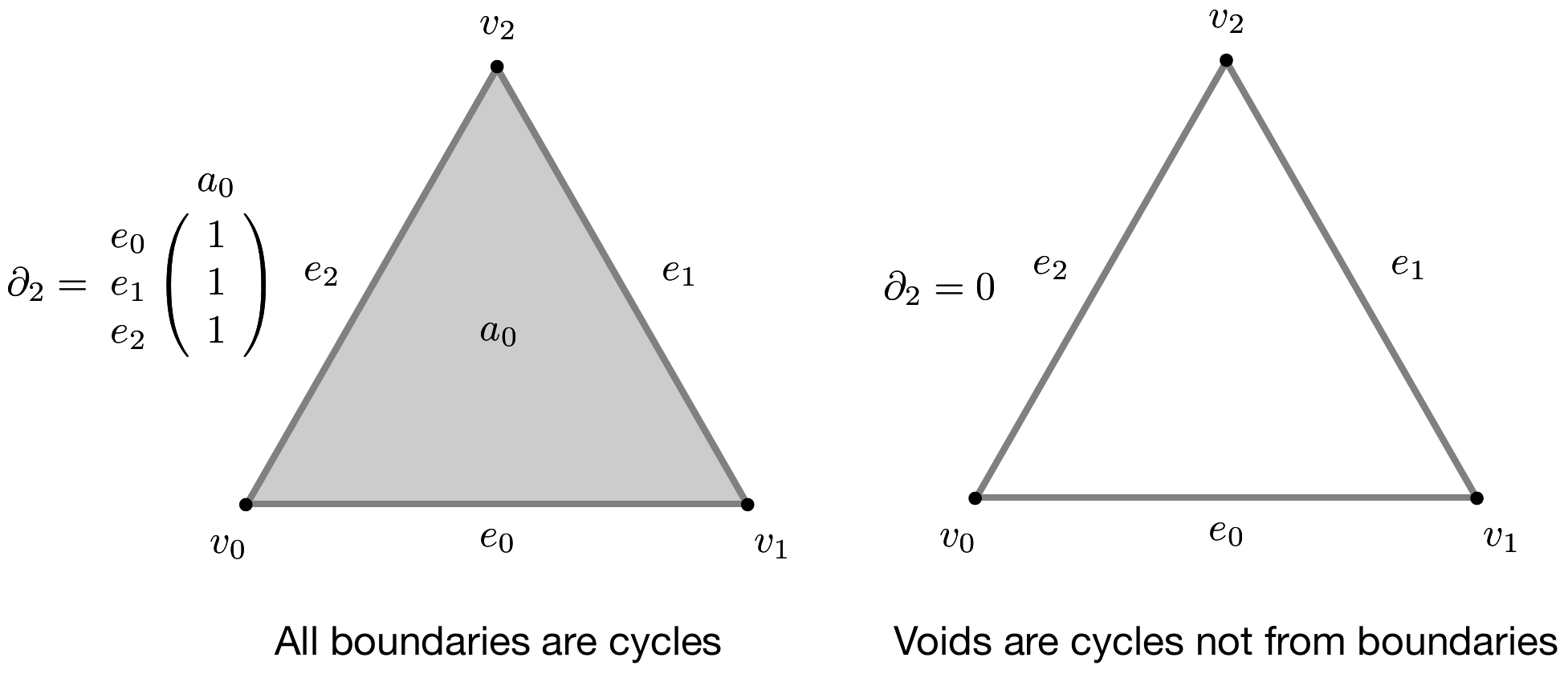}
    \caption{The boundary matrix $\partial_2$ is capturing if a cycle is a void or the boundary of a filling simplex?}
    \label{fig:homology2}
\end{figure}

We are however not done computing homology just by computing the number of cycles. After all, the very example we have showed a filling area $a_0$, hence the cycle is actually filled, hence does not bound a void. This is precisely the condition we are missing. We need to be able to differentiate if a cycle is filled in or not. The condition is straightforward. Recall that each $n$-simplex by definition contains cycles that bound it. This information is encoded in the next higher-dimensional boundary matrix. So let us construct the $\partial_2$ boundary matrix for both possible cases. A cycle of three edges is either bounding a void or is the boundary of an area (or in the simplicial language a $2$-simplex). Both cases are shown in Figure \ref{fig:homology2}.

\begin{figure}[h]
    \centering
    \includegraphics[width=.95\textwidth]{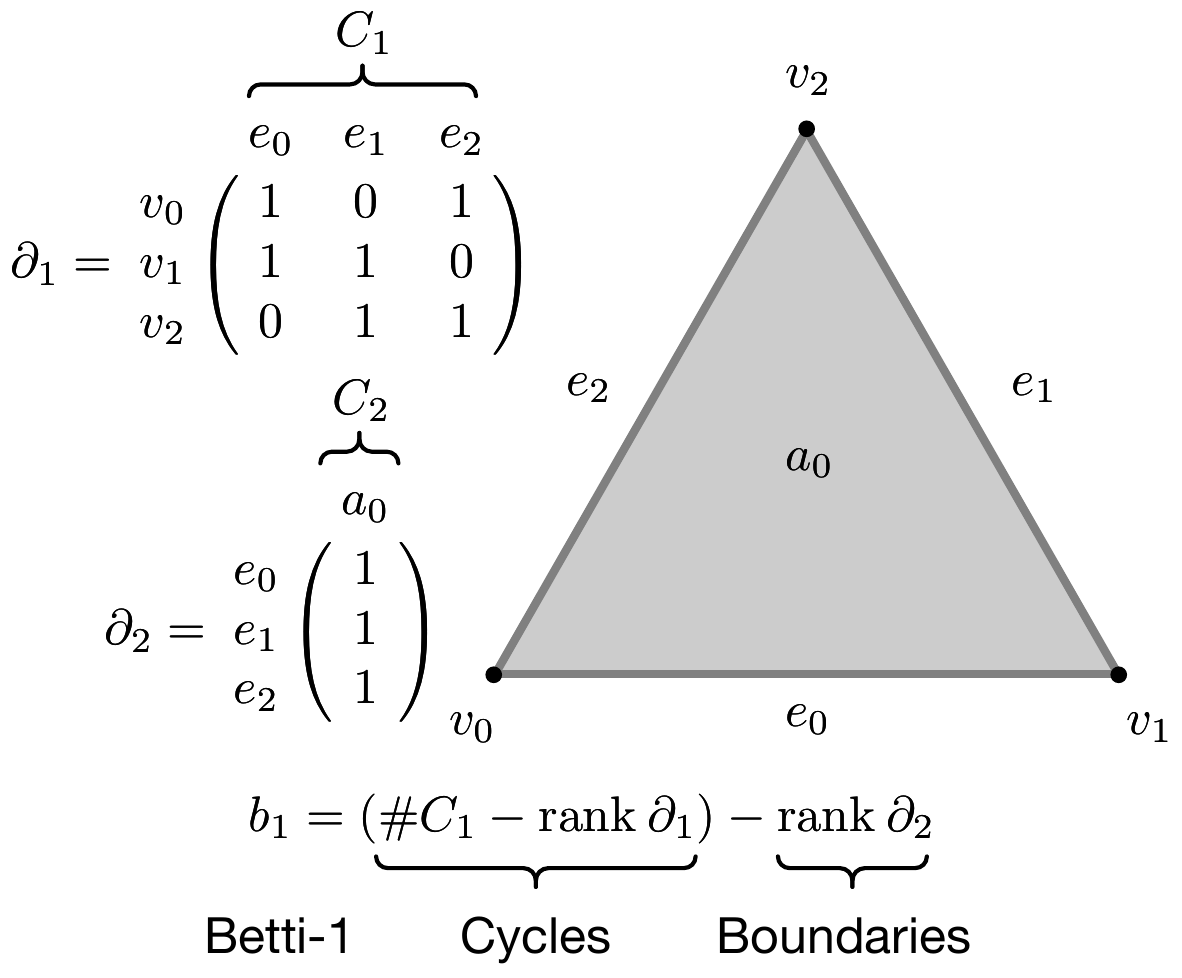}
    \caption{Computation of Betti-1 $b_1$, the number of area-like voids.}
    \label{fig:Betti1}
\end{figure}

\begin{figure}[h]
    \centering
    \includegraphics[width=.65\textwidth]{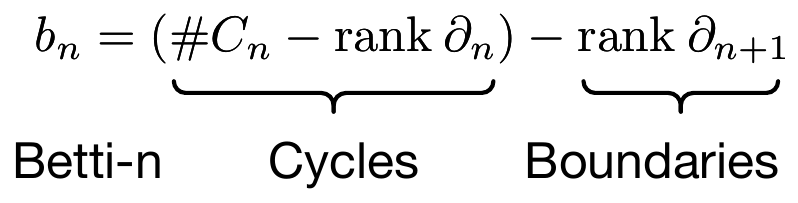}
    \caption{Computation of Betti-N $b_n$, the number of n-simplex-like voids.}
    \label{fig:Bettin}
\end{figure}

If the area is filled then the boundary map contains a map from the area to the three edges in its boundary. Given that we only have one area in this example, we get only one column in the boundary matrix $\partial_2$. If, however, the cycle does not come from the boundary of an area, there is no matching map in the boundary map. In other words, cycles generated from being a boundary require a linearily independent entry in the boundary matrix, hence the rank must be containing all of them.
Now we have all the pieces to compute homology. The information if something is in the cycle is in the lower dimensional boundary matrix, specifically in the size of its null space. The information if something is filled in or not is contained in the higher dimensional boundary matrix, specifically in the rank of the boundary matrix.
All this is collected in this final Figure \ref{fig:Betti1} on the computation of the simplicial homology.
It turns out that these relationships hold for any dimensional simplicial complexes and their boundary matrices derived analogous to how we did it for $1$- and $2$-simplices. So we get the formula for Betti numbers as shown in Figure \ref{fig:Bettin}. 
This, in a nutshell, is the computation of homology of a simplicial complex! It amounts to computing the rank (and by simple extension the size of the null-space) of two matrices. There are a plethora of ways to compute the rank of a matrix, and variations of Gaussian elimination should come to mind. A particular form of a matrix, the {\em Smith Normal Form} can be computed by a Gaussian elimination style reduction and it makes reading off rank easy. A point to keep in mind here is that the specifics of the columns of the matrix does not matter for these results, merely that they are linearily (in)dependent as we discussed before.
\subsection{Homology Groups}
What we have discussed so far is how one computes homology for a simplicial complex. However, virtually all modern textbooks on algebraic topology will describe homology through groups. To appreciate the connection of our discussion above in terms of constructed boundary matrices and rank-nullity computations, here is a brief peek at the algebraic formulation of the same ideas.

The notion of an $n$-Chain $C_n$ captures our process of constructing boundary matrices and what they operate upon. Chains that are connected via a sequence of boundary operators are called a {\em Chain complex}.

\begin{figure}
\begin{equation*}
    0\xrightarrow{}C_n\xrightarrow{\partial_n}C_{n-1}\xrightarrow{\partial_{n-1}}\cdots\xrightarrow{} C_1\xrightarrow{\partial_{1}}C_0\xrightarrow{\partial_{0}}0
\end{equation*}

\centering
\begin{tikzpicture}[
node distance = 0 mm and 33mm,
     E/.style = {shape=ellipse, aspect=0.7,
                 minimum height=2mm+#1mm,
                 minimum width=#1mm,
                 draw, anchor=south,
                 node contents={}}
                    ]
    \node (n3a) [fill=black!5,E={9+21}];
    \node (n2a) [fill=black!10,E={9+14}];
    \node (n1a) [fill=black!15,E={9+7}];
    \node (n0a) [draw,fill=white,shape=circle,inner sep=0,minimum size =3pt] {};
    
    \node (m1a) [minimum width=3em,below=of n1a.north] {$\boldsymbol{B}_{n+1}$};
    \node (m2a) [minimum width=3em,below=of n2a.north] {$\boldsymbol{Z}_{n+1}$};
    \node (m3a) [minimum width=3em,below=of n3a.north] {$\boldsymbol{C}_{n+1}$};
    \node (m0a) [minimum width=3em,below=of n0a.south] {$0$};


    \node (n3b) [right=30mm,fill=black!5,E={9+21}];
    \node (n2b) [right=30mm,fill=black!10,E={9+14}];
    \node (n1b) [right=30mm,fill=black!15,E={9+7}];
    \node (n0b) [right=30mm,draw,fill=white,shape=circle,inner sep=0,minimum size =3pt] {};

    \node (m1b) [minimum width=3em,below=of n1b.north] {$\boldsymbol{B}_{n}$};
    \node (m2b) [minimum width=3em,below=of n2b.north] {$\boldsymbol{Z}_{n}$};
    \node (m3b) [minimum width=3em,below=of n3b.north] {$\boldsymbol{C}_{n}$};
    \node (m0b) [minimum width=3em,below=of n0b.south] {$0$};
    
    \draw[black] (n3a.north) .. controls +(right:10mm) and +(left:10mm) .. (n1b.north);
    \draw[black] (n2a.north) .. controls +(right:10mm) and +(left:10mm) .. (n1b.south);

    \draw[black] (n0b) -- node[midway, below] {$\xrightarrow[\partial_{n+1}]{\qquad}$} (n0a);

    \node (n3c) [right=60mm,fill=black!5,E={9+21}];
    \node (n2c) [right=60mm,fill=black!10,E={9+14}];
    \node (n1c) [right=60mm,fill=black!15,E={9+7}];
    \node (n0c) [right=60mm,draw,fill=white,shape=circle,inner sep=0,minimum size =3pt] {};

    \node (m1c) [minimum width=3em,below=of n1c.north] {$\boldsymbol{B}_{n-1}$};
    \node (m2c) [minimum width=3em,below=of n2c.north] {$\boldsymbol{Z}_{n-1}$};
    \node (m3c) [minimum width=3em,below=of n3c.north] {$\boldsymbol{C}_{n-1}$};
    \node (m0c) [minimum width=3em,below=of n0c.south] {$0$};

    \draw[black] (n3b.north) .. controls +(right:10mm) and +(left:10mm) .. (n1c.north);
    \draw[black] (n2b.north) .. controls +(right:10mm) and +(left:10mm) .. (n1c.south);
    \draw[black] (n0c) -- node[midway, below] {$\xrightarrow[\partial_{n}]{\qquad}$} (n0b);
%
\end{tikzpicture}

\begin{equation*}
    H_n=Z_n/B_n=\ker \partial_n/\Ima \partial_{n+1}
\end{equation*}
\caption{Homology in the language of abelian groups and $n$-chains. The sequence of chains is called a chain complex.}
\end{figure}
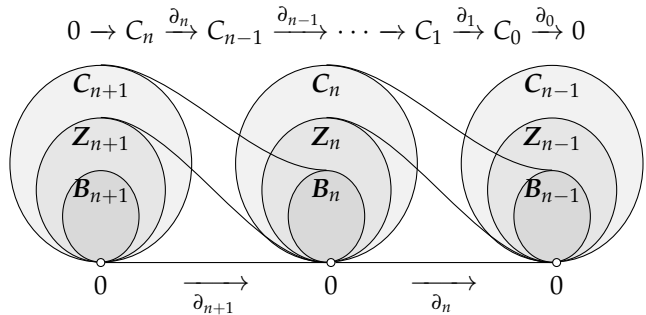
As we have seen, some subset of chains can form cycles, which here is captured by $Z_n$. Furthermore, some chains are present because they are in the boundary of some higher dimensional element in a $C_{n+1}$ chain, and therefore are in the {\em image} (or $\im$) of the boundary map from $\partial_{n+1}:C_{n+1}\xrightarrow{} C_{n}$. Furthermore, recall that we noted that cycles $Z_n$ are characterized by falling into the null-space which is also known as the {kernel} (or $\ker$) of the boundary map $\partial_n$. Homology are the voids that are not from boundaries, hence we get the formula for the $n$th Homology group that is the cycles in the chain $Z_n$ with the boundaries $B_n$ i the chain "modded out" (that is removed), and we see that we can compute this information from the kernel and image of two boundary maps as we saw before.

Finally, observe that the boundary of a boundary must always be zero, given that any boundary chains $B_n$ are fully included in the cycles $Z_n$ and all cycles are send to $0$ by the second boundary map. This fact is called the {\em fundamental lemma of Homology}.
\begin{figure}
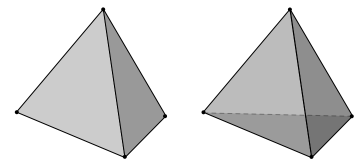

\begin{equation*}
\partial_{n}\partial_{n+1}=0
\end{equation*}
\caption{The fundamental lemma of Homology states that a boundary does itself have no boundary.}
\end{figure}

\begin{marginfigure}
    \centering
    \includegraphics[width=.95\marginparwidth]{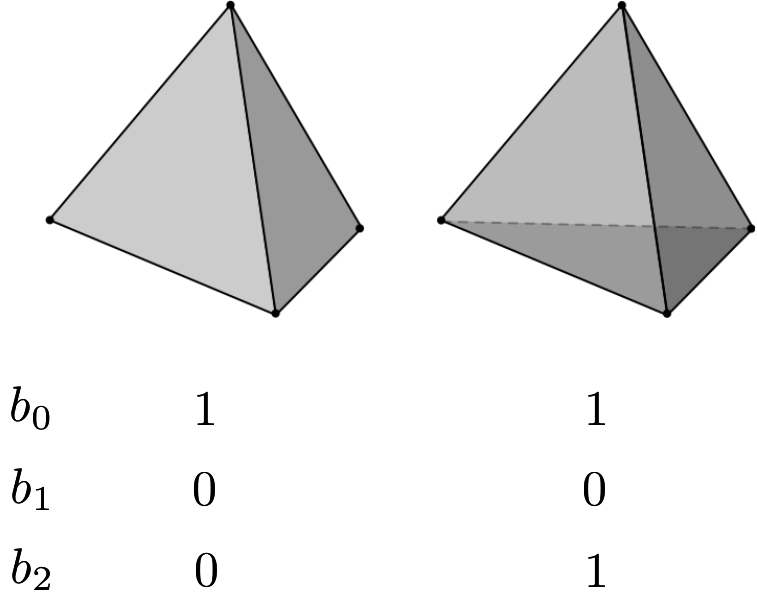}
    \caption{The Betti numbers of the simplicial ball and sphere.}
    \label{fig:Bettiex1}
\end{marginfigure}

\begin{marginfigure}
    \centering
    \includegraphics[width=.95\marginparwidth]{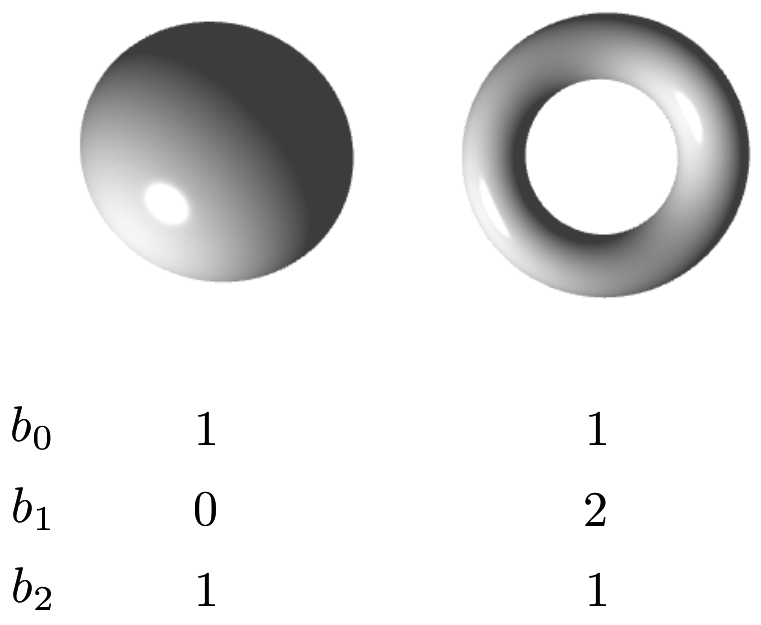}
    \caption{The Betti numbers of the ball and torus.}
    \label{fig:Bettiex2}
\end{marginfigure}

To get a sense of what homology looks like in concrete cases, here are some simple examples: Figure \ref{fig:Bettiex1} shows a solid tetrahedron (that you can also think of as an abstract $3$-simplex) and next to it the boundary of said shape with the interior empty. We see that both cases are a single connected component ($b_0=1$), while only the hollow shape has $b_2=0$. While computationally we will generally deal with simplicial complexes, remember that conceptually we are dealing with deformability, and if we inflate our hollow tetrahedron in the physical world we will arrive at something that looks like the surface of a sphere as seen in Figure \ref{fig:Bettiex2}. Next to it we depict the torus and its Betti numbers. The torus is characterized by two loops that cannot be collapsed as we have seen before and this is captured here by $b_1=2$. 
Sometimes homological information is not presented in terms of Betti numbers but rather in terms of the structure of the Homology group. In our case, the groups are finitely generated. The cycles (without bindaries) are the generators in this group. Hence we can draw upon the structure theorem for finitely-generated abelian groups with respect to direct sums which is as follows:
\begin{thm}
A {\bf finitely-generated abelian group} $A$ can be uniquely expressed in the form of direct sums of finite numbers of free cyclic groups $\mathbb{Z}$, and cyclic groups $\mathbb{Z}_{t_i}$ of finite period $t_I$ called {\bf Torsion}. The rank of the free group we will call {\bf Betti numbers}, The indices $t_i$ are not necessarily distinct prime and are called {\bf torsion coefficients}:
\begin{equation*}
A = \bigoplus\limits_{b}\mathbb{Z}\oplus \bigoplus\limits_{i}\mathbb{Z}_{t_i}
\end{equation*}
\end{thm}
\begin{marginfigure}
    \centering
    \includegraphics[width=.95\marginparwidth]{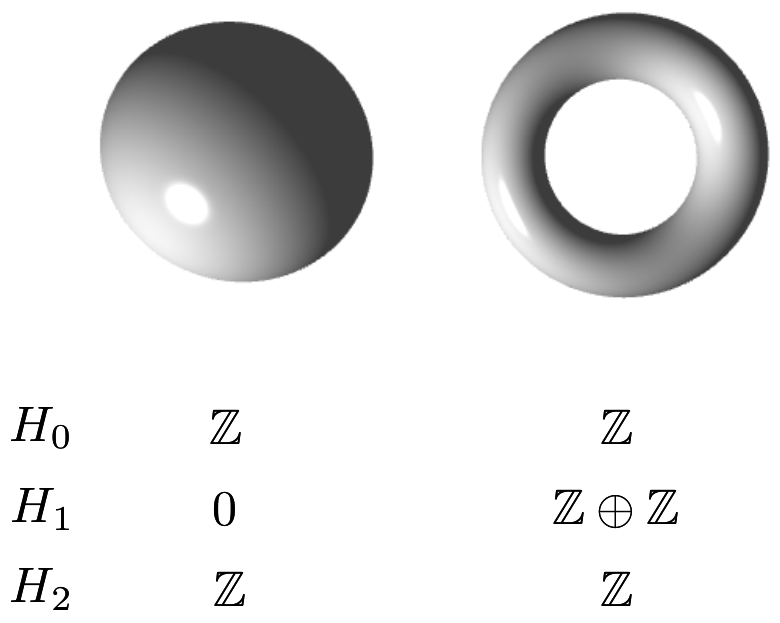}
    \caption{The homology groups of the ball and torus.}
    \label{fig:Bettiex3}
\end{marginfigure}
In this view, Betti numbers count the number of free groups $\mathbb{Z}$ and hence the homology groups for the sphere and torus written as shown in Figure \ref{fig:Bettiex3}. In our discussion we have ignored the {\em Torsion} coefficients that show up in this structure theorem. If we use our xor linear algebra, torsion never occurs, but it can occur if we consider oriented situations. However, so far torsion coefficients have not made a particular impression in applied topological examples, so we will continue to ignore them in this introduction. But it is good to know they exist!

\subsection{Smith Normal Form}

\begin{figure}[h]
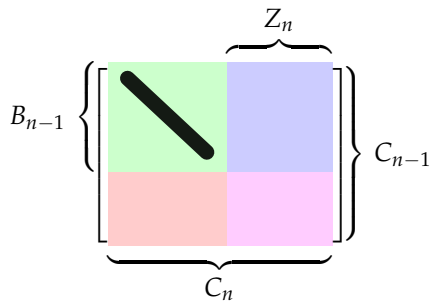

\[\begin{bNiceMatrix}[first-col,last-col,margin]
   &\Block[fill=[RGB]{204,255,204}]{3-3}{}\phantom{1} &  & & \Block[fill=[RGB]{204,204,255}]{3-3}{} & & \\
B_{n-1}\phantom{II}   & & \phantom{1} &  & & & & \\
   & &  & \phantom{1} & & & & \phantom{II}C_{n-1} \\
   & \Block[fill=[RGB]{255,204,204}]{2-3}{} &  &  & \Block[fill=[RGB]{255,204,255}]{2-3}{} \phantom{1} & & & \\
   & & & & & \phantom{1} & & \\
\CodeAfter
  \tikz \draw [opacity=.9,line width=2mm,line cap=round] 
              (1-1.center) -- (3-3.center);
      \OverBrace[yshift=5pt]{1-4}{1-6}{Z_n}
      \UnderBrace[yshift=5pt]{5-1}{5-6}{C_n}
      \SubMatrix{\{}{1-1}{3-2}{.}[left-xshift=8pt]
      \SubMatrix{.}{1-2}{5-5}{\}}[xshift=18pt]
\end{bNiceMatrix}
\]
\vspace{8pt}
\caption{Rank information captured in the boundary map related to $C_n$ and $C_{n-1}$ in Smith Normal Form.}\label{fig:snf}
\end{figure}

Given that the one key piece of information to compute homology is the rank of two matrices, ultimately one wants algorithms that serve that purpose. This itself is a substantial topic on its own right. Instead of trying to do the right thing and cover this fully, we will limit ourselves to hint at the structure in which one can see rank information clearly. In a general matrix rank is not easily inspected. However, some matrix forms will make rank easy or even trivial to inspect. One important such form is the {\em Smith Normal Form} which is depicted in Figure \ref{fig:snf}.

\noindent The Smith Normal Form has the form of having a non-zero entry in the diagonal. This captures a particularly nice basis and makes explicit the rank, which is precisely the number of rows/columns were one can find these diagonal entries. Observe that the image of the boundary information of the matrix $B_{n-1}$ is precisely the size of this diagonal block that is indeed square. This will be important to remember later, as this shape will therefore not change as you transpose! The kernel of the matrix $Z_n$ is not necessarily square in shape and in fact can disappear if the matrix is max rank. Its shape is in fact not the same under transpose.

The Smith Normal Form is computed by row or column reductions akin to Gaussian elimination. Given that this is central to fast homology computation, fast algorithms have been developed to compute it quickly. However, for exploration and understanding, straight-forward naive implementation is very helpful to study!

\subsection{Homology of Graphs}

Given the importance of graphs, we want to study the homology of graphs. Homology is not often exposed relative to graph theory, though this has changed in recent years. To start off, consider the chain complex of a graph:

\[0\xrightarrow[]{\partial_2} C_1 \xrightarrow[]{\partial_1} C_0 \xrightarrow[]{\partial_0}0 \]

There is only one boundary map $\partial_1$ as only the edge-vertex relationship contains non-trivial information. The map $\partial_0$ goes to zero. This means that all vertices are considered cycles! We will see the consequence of this choice when we discuss connected component computation. If this map is not zero the associated homology is related to a notion of {\em reduced homology} which we will not cover here. The second map to note is $\partial_2$ whose image contains no information, hence there are no boundaries from $2-$-simplices in graphs, as one should expect.

There are two positions for the fundamental lemma of homology:
\[\partial_1\partial_{2}=0\qquad\partial_0\partial_1=0\] 
which tells us that indeed there are two dimensions in which we get interesting homology, $b_1$ which from our previous discussion we already know counts the number of cycles in the graph\marginnote{$b_1$ counts the number of cycles and $b_0$ counts the number of connected components in a graph.}  and $b_0$ which counts the number of connected components. Why is that? 
\begin{figure}
     \centering
    \includegraphics[width=.95\textwidth]{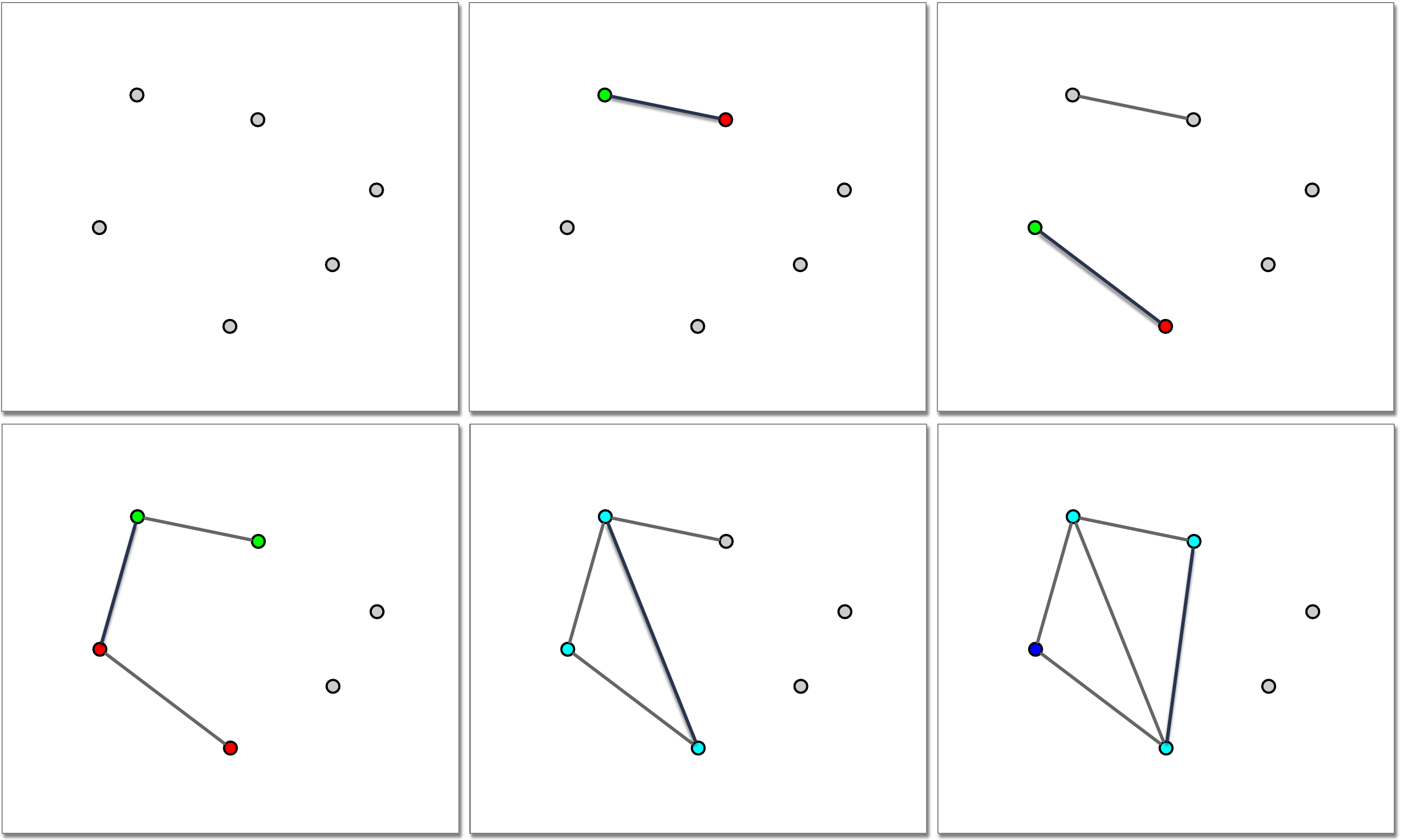}
    \caption{Computing the connected component of a graph.}
    \label{fig:connectedcomponent}
\end{figure}
Consider Figure \ref{fig:connectedcomponent}. In it we demonstrate the homology computation for $b_0$ by starting off with a set of six isolated vertices. Given that none are connected each vertex is its own connected component. Hence we would expect the homology computation to give us $b_0=6$. Given that all vertices are cycles ($\ker \partial_0$ is true for all of $\partial_0$, i.e. all vertices). Notice that this fact never changes. So, for every configuration, it will now always count the number of vertices as cycles. Then we add an edge, this reduces the number of connected components as two vertices are now connected. The new correct answer is now $b_0=5$. In our homology computation the edge shows up in the map $\Ima\partial_1$ hence we get one rank from that matrix and we get $b_0=6-1=5$ which is the expected result. the same happens for the next to steps in our figure, and we observe that it does not matter if edges connect vertices or larger connected compontents. However, then we add an edge, which gives us a new case in the bottom middle of our cartoon. An edge is added, but we observe that the number of connected components we can count did not decrease. Let us use the homology computation for a hint. The new edge actually forms a cycle! Hence it does not create more rank in the boundary matrix $\partial_1$. Therefore $b_0=6-3=3$ despite there being four edges. This checks out as the edge only connected vertices that were already previously connected. The final picture give another edge that does not change the number of connected components. We now have all the cases for an inductive proof that $b_0$ computes connected components. Note that this argument is valid even for higher-order simplicial complexes. $b_0$ always gives us the number of connected components for any simplicial complex.

\section{Cohomology}

So far we have used boundary matrices and face maps in all our constructions. This is the same thing as moving down in dimensions as the boundary matrix $\partial_n: C_n\rightarrow C_{n-1}$. But this is just a choice. We could just as well decide to move up in dimensions! In modern mathematics the prefix {\em co-} is associated with flipping directions of maps around.

We have notated our boundary matrices in such a way that the input is associated with columns and the output is associated with rows. So a simple formal way to change the direction in which we are going is to swap inputs and output. Here we show how this looks for the example of the 2-simplex we discussed before:

\begin{figure}
\[
\partial_1=\bordermatrix{
  ~ & e_0 & e_1 & e_2 \cr
  v_0 & 1 & 0 & 1 \cr
  v_1 & 1 & 1 & 0\cr
  v_2 & 0 & 1 & 1\cr
} \xleftrightarrow[\text{transpose}]{\text{"co"}}\partial^T_1=\bordermatrix{
  ~ & v_0 & v_1 & v_2 \cr
  e_0 & 1 & 1 & 0 \cr
  e_1 & 0 & 1 & 1\cr
  e_2 & 1 & 0 & 1\cr
} 
\]

\[
\partial_2=\bordermatrix{
  ~ & a_0 \cr
  e_0 & 1 \cr
  e_1 & 1 \cr
  e_2 & 1 \cr
} \xleftrightarrow[\text{transpose}]{\text{"co"}}\partial_2^T=\bordermatrix{
  ~ & e_0 & e_1 & e_2\cr
  a_0 & 1 & 1 & 1\cr
}
\]
\caption{The matrix transpose swaps inputs and outputs hence changes directions of the map it represents. The flipping of the map is associated with adding or removing the "co-" prefix.}
\end{figure}

Hence it makes sense to notate the transpose boundary matrix when we simply flipped a boundary map from going down in dimensions to going up. However, it is customary to index maps with respect to the dimension of the tail of the arrow and we see that flipping the arrow actually placed $C_{n-1}$ at the tail. One can find an additional notation in the literature that uses the $\delta_n$ notation which just shifted the dimension index. This new operator is called the {\em coboundary matrix}\marginnote{$\delta_n=\partial^T_{n+1}$ is the coboundary matrix.}.
$$\partial_n: C_n\rightarrow C_{n-1}$$
$$\partial_n^T: C_n\leftarrow C_{n-1} \qquad \delta_n: C^{n+1}\leftarrow C^{n}$$

Notice that we have moved from subscripts to superscripts for the coboundary map. It is a widely-used convention to denote homological maps and structures with a subscript, and cohomology maps and structures with a superscript. This hints at things going down in dimension in homology, and up in dimension in cohomology. Also when you encounter a superscript the entity is the "co" version. For example $Z_n$ are cycles, while $Z^N$ are cocycles! In most of these notes we will stay homological, and we will use the notation $\partial^T_n$ to indicate boundary maps that have been transposed to go up to indicate that we generally keep homology as our starting perspective.

\begin{figure}
\begin{equation*}
    0\xleftarrow{\delta_n}C^n\xleftarrow{\delta_{n-1}}C^{n-1}\xleftarrow{\delta_{n-2}}\cdots\xleftarrow{} C^1\xleftarrow{\delta_{0}}C^0\xleftarrow{}0
\end{equation*}

\centering
\begin{tikzpicture}[
node distance = 0 mm and 33mm,
     E/.style = {shape=ellipse, aspect=0.7,
                 minimum height=2mm+#1mm,
                 minimum width=#1mm,
                 draw, anchor=south,
                 node contents={}}
                    ]
    \node (n3a) [fill=black!5,E={9+21}];
    \node (n2a) [fill=black!10,E={9+14}];
    \node (n1a) [fill=black!15,E={9+7}];
    \node (n0a) [draw,fill=white,shape=circle,inner sep=0,minimum size =3pt] {};
    
    \node (m1a) [minimum width=3em,below=of n1a.north] {$\boldsymbol{B}^{n+1}$};
    \node (m2a) [minimum width=3em,below=of n2a.north] {$\boldsymbol{Z}^{n+1}$};
    \node (m3a) [minimum width=3em,below=of n3a.north] {$\boldsymbol{C}^{n+1}$};
    \node (m0a) [minimum width=3em,below=of n0a.south] {$0$};


    \node (n3b) [right=30mm,fill=black!5,E={9+21}];
    \node (n2b) [right=30mm,fill=black!10,E={9+14}];
    \node (n1b) [right=30mm,fill=black!15,E={9+7}];
    \node (n0b) [right=30mm,draw,fill=white,shape=circle,inner sep=0,minimum size =3pt] {};

    \node (m1b) [minimum width=3em,below=of n1b.north] {$\boldsymbol{B}^{n}$};
    \node (m2b) [minimum width=3em,below=of n2b.north] {$\boldsymbol{Z}^{n}$};
    \node (m3b) [minimum width=3em,below=of n3b.north] {$\boldsymbol{C}^{n}$};
    \node (m0b) [minimum width=3em,below=of n0b.south] {$0$};
    
    \draw[black] (n3b.north) .. controls +(left:10mm) and +(right:10mm) .. (n1a.north);
    \draw[black] (n2b.north) .. controls +(left:10mm) and +(right:10mm) .. (n0a);

    \draw[black] (n0a) -- node[midway, below] {$\xleftarrow[\delta_{n}]{\qquad}$} (n0b);

    \node (n3c) [right=60mm,fill=black!5,E={9+21}];
    \node (n2c) [right=60mm,fill=black!10,E={9+14}];
    \node (n1c) [right=60mm,fill=black!15,E={9+7}];
    \node (n0c) [right=60mm,draw,fill=white,shape=circle,inner sep=0,minimum size =3pt] {};

    \node (m1c) [minimum width=3em,below=of n1c.north] {$\boldsymbol{B}^{n-1}$};
    \node (m2c) [minimum width=3em,below=of n2c.north] {$\boldsymbol{Z}^{n-1}$};
    \node (m3c) [minimum width=3em,below=of n3c.north] {$\boldsymbol{C}^{n-1}$};
    \node (m0c) [minimum width=3em,below=of n0c.south] {$0$};

    \draw[black] (n3c.north) .. controls +(left:10mm) and +(right:10mm) .. (n1b.north);
    \draw[black] (n2c.north) .. controls +(left:10mm) and +(right:10mm) .. (n0b);
    \draw[black] (n0b) -- node[midway, below] {$\xleftarrow[\delta_{n-1}]{\qquad}$} (n0c);
%
\end{tikzpicture}

\begin{equation*}
    H^n=Z^n/B^n=\ker \delta_n/\Ima \delta_{n-1}
\end{equation*}
\caption{Cohomology in a nutshell.}\label{fig:cohomology}
\end{figure}
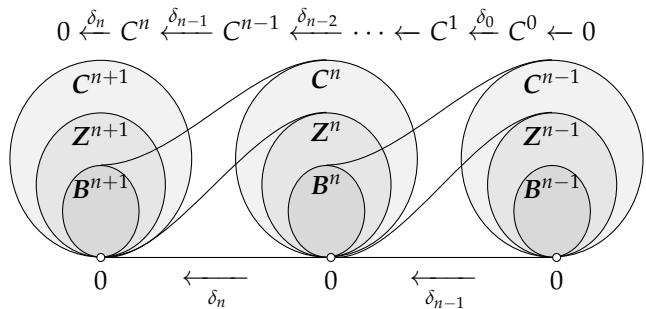

Cohomology in this view is just homology with maps flipped. Hence the egg diagram we used before for homology now looks as follows (Figure \ref{fig:cohomology}).
The key observation here is that we can still compute homological information as the quotient of the kernel of one map and the image of another map! But how we grab this information has changed. We are getting the image from below and the kernel from above. This means that the direct intuition of boundary cycles is no longer correct. Formally, we are computing the quotient of the Cocycle with the Coboundary. In our context the Betti number so computed will agree with those from homology, so cohomology is just a differently (dually) organized homology theory.

\subsection{Orientation and Coefficients}

Up to this point we have operated only with entries of $0$ and $1$ in our boundary matrix to capture connectivity information. But we might want to capture more information. For example we may want to keep orientation information. A single digit binary number is not enough to capture this.

\begin{figure}[h]
    \centering
    \includegraphics[width=.95\textwidth]{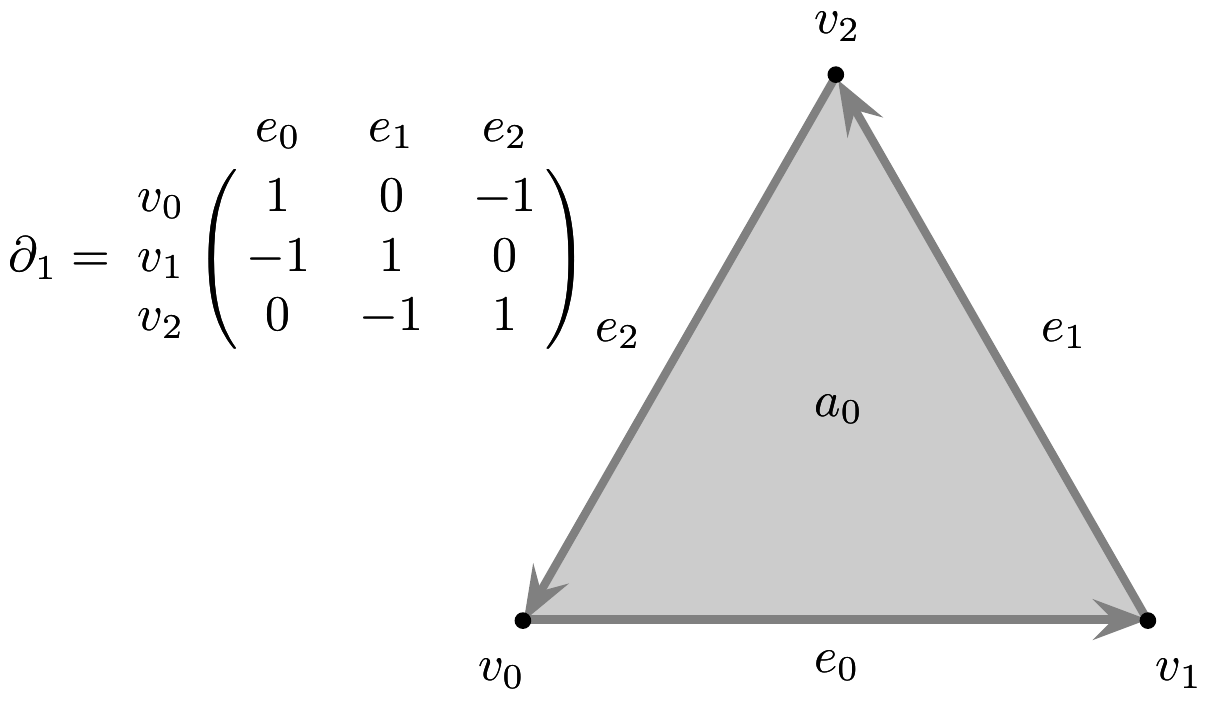}
    \caption{Boundary matrix of oriented $1$-simplices. The addition of all three columns again leads to a zero vector.}
    \label{fig:orientedhomology}
\end{figure}

First let us check that our linear dependence argument actually works with orientation. To this end we will use signed numbers, and the convention that the base of an arrow is $1$ and the tip of the arrow is $-1$ to construct a boundary matrix as shown in Figure \ref{fig:orientedhomology}. It is easy to check that again the sum of the three columns equals the zero vector. So the argument that oriented cycles create a linear dependency in the boundary matrix works out as before.
However, rather than just sticking with binary plus sign, we will immediately expand the coefficients we allow in our constructions to be in $\mathbb{R}$. While (co)homology can be done with other coefficients, such as rings or cyclic groups, this leads to some extra complications that we are happy to avoid in this discussion. The interested reader is invited to study the {\em universal coefficient theorem}\marginnote{Working in $\mathbb{R}$ allows us to avoid complications caused by torsion, as characterized by the universal coefficient theorem.}. It turns out that using binary coefficients or coefficients in $\mathbb{R}$ torsion effects (which essentially cause the complication) disappear. Luckily $\mathbb{R}$ is a rather typical setting in many applications so what we would tend to pick coincides with an easier setting!

\subsection{Inner Product in Matrix Multiplication}

So far we really have used linear algebra to characterize rank as a way to capture linear dependence as a way to capture when a boundary closes on itself. Perhaps we could have done the same thing with logic and replaced our $0$ and $1$ with {\em false} (not a boundary) and {\em true} (a boundary). But we have just allowed ourselves to use coefficients for much more than just a binary state of connectivity. This gives us a pathway to expand that amount of linear algebra we are using, and specifically give these coefficient meaning beyond just connectivity.

One interesting operation to allow oneself to use is matrix multiplication as it allows us to compose the boundary maps we have constructed. Furthermore we have seen that the transpose allows us to change the direction we are going. The standard matrix multiplication combines the row dimensions of the first matrix with matching column dimensions of the second matrix. Notice that our transposes always create matching dimensions! So we know we can compute matrix multiplications between a matrix and its transpose, and this work in either order. This process is shown in Figure \ref{fig:matrixmultiplication}.

\begin{figure}
\begin{align*}
\begin{pNiceMatrix}[first-row,margin]
\Block[]{1-4}{\partial^T}\\
\phantom{-}1 & -1 & \phantom{-}0 & \phantom{-}0 \\
\Block[fill=red!30,rounded-corners]{1-4}{}
\phantom{-}0 & \phantom{-}1 & -1 & \phantom{-}0 \\
\phantom{-}0 & \phantom{-}0 & \phantom{-}1 & -1 \\
\end{pNiceMatrix}\cdot
\begin{pNiceMatrix}[first-row,margin]
\Block[]{1-3}{\partial}\\
\phantom{-}1 & \phantom{-}0 & \Block[fill=blue!15,rounded-corners]{4-1}{}\phantom{-}0\\
-1 & \phantom{-}1 & \phantom{-}0\\
\phantom{-}0 & -1 & \phantom{-}1\\
\phantom{-}0 & \phantom{-}0 & -1\\
\end{pNiceMatrix}&=
\begin{pNiceMatrix}[first-row,margin]
\Block[]{1-3}{L^{\text{up}}}\\
\phantom{-}2 & -1 & \phantom{-}0\\
-1 & \phantom{-}2 & \Block[fill=green!30,rounded-corners]{1-1}{}-1\\
\phantom{-}0 & -1 & \phantom{-}2\\
\end{pNiceMatrix}\\
\begin{pNiceMatrix}[first-row,margin]
\Block[]{1-3}{\partial}\\
\phantom{-}1 & \phantom{-}0 & \phantom{-}0\\
\Block[fill=red!30,rounded-corners]{1-3}{}
-1 & \phantom{-}1 & \phantom{-}0\\
\phantom{-}0 & -1 & \phantom{-}1\\
\phantom{-}0 & \phantom{-}0 & -1\\
\end{pNiceMatrix}\cdot
\begin{pNiceMatrix}[first-row,margin]
\Block[]{1-4}{\partial^T}\\
\phantom{-}1 & -1 & \Block[fill=blue!15,rounded-corners]{3-1}{} \phantom{-}0 & \phantom{-}0 \\
\phantom{-}0 & \phantom{-}1 & -1 & \phantom{-}0 \\
\phantom{-}0 & \phantom{-}0 & \phantom{-}1 & -1 \\
\end{pNiceMatrix}&=
\begin{pNiceMatrix}[first-row,margin]
\Block[]{1-4}{L^{\text{down}}}\\
\phantom{-}1 & -1 & \phantom{-}0 & \phantom{-}0\\
-1 & \phantom{-}2 & \Block[fill=green!30,rounded-corners]{1-1}{}-1 & \phantom{-}0\\
\phantom{-}0 & -1 & \phantom{-}2 & -1\\
\phantom{-}0 & \phantom{-}0 & -1 & \phantom{-}1\\
\end{pNiceMatrix}
\end{align*}
\caption{The matrix multiplication of the boundary matrix with its transpose in both possible orders. Observe that the computation of each outcome has the structure of a dot or inner product.}\label{fig:matrixmultiplication}
\end{figure}

In color it indicates how a matrix multiplication actually collects information to compute a single resulting number. This is a dot or inner product. Hence we are actually using an inner product when we perform a matrix multiplication.

\subsection{Inner product}

Usually one just uses what is called the {\em standard inner product} in linear algebra, but we want to be a little bit more detailed here. An inner product defines a map from a pair of vectors to a number and the number measures the projection of one vector onto the other (this is symmetric!). However this number is in general not unique. We can decide to weigh the outcome. Consider two definition of an inner product in Figure \ref{fig:innerproduct}:

\begin{figure}
\[\langle x,y\rangle_{C_n}=\sum_{i=0}^{\#C_n-1}x_i\cdot y_i\qquad\langle x,y\rangle^{w}_{C_n}=\sum_{i=0}^{\#C_n-1} w_i\cdot x_i\cdot y_i\]
\caption{The standard and the weighted inner product.}\label{fig:innerproduct}
\end{figure}

The standard inner product is really just the inner product with scalar weights where all weights are $1$. The inner product introduces some geometry because we can now measure how much of a vector projects onto another vector by some number, and the weight gives us some control over how strongly each dimension contributes. This is important! By using the inner product we are no longer just doing topology, we are picking up some local geometry as well. This is conceptually very important because it allows us to be clear about what aspects come from topology (essentially everything we can do without the inner product) and those that are geometrical (everything that necessarily requires the inner product). Typically everything geometrical has a topology (albeit it can be a very boring one) and this is captured here. We are still using the boundary matrices as before but they additionally now have some metric information. What we mean by "geometry" is, however, complicated. We created a choice of weights. What need or want these weights address  usually depends on the problem space we are investigating.

Given that all the inner products we will encounter are over a vector space related to an $n$-chain, we include this in our inner product notation. Also notice that in our matrix multiplication example the length of the vectors actually differs depending on the order in which a boundary matrix and its transpose are multiplied. We really need two definitions of inner products, one over each $n$-chain involved:

\[\langle,\rangle_{C_{n+1}}\qquad\langle,\rangle_{C_n}\]

\subsection{The Adjoint}

When discussing cohomology we discussed the transpose as flipping inputs and outputs. With coefficients something more can happen, given that we are now involving an inner product structure.

The process to define the inverting of relating the boundary matrix to a coboundary matrix can now be constructed through the inner product.

\begin{figure}
\[\langle \partial x_{n+1},x_{n}\rangle_{C_{n}} = \langle x_{n+1},\partial^* x_{n}\rangle_{C_{n+1}} \]
\[x_{n+1}\in C_{n+1},\quad x_n\in C_{n}\]
\caption{Definition of the adjoint $\partial^*$. It relates boundary and coboundary matrix through the inner product. We want our inner product to agree between dimensions.}
\end{figure}

With coefficients in $\mathbb{R}$ and the standard inner product, the adjoint $\partial^*$ turns out to be just the transpose $\partial^T$ as before. Hence we will use the notation $\partial^T$ throughout the rest of the notes, though in other exposition you may encounter the adjoint notation.

Given that the inner product wants vectors from the same vector space, both arguments have to be in it. Notice that  $x_{n+1}$ is not in $C_n$ but it is when going through the boundary matrix $\partial_{n+1}$. So both arguments are indeed in $C_n$ as required. We want our metric information to agree between dimensions. Hence the equality. Then if we take the coboundary matrix $\partial^*$ on a vector $x_n$ which sends it to $C{n+1}$ and we inner product it with a vector $x_{n+1}$ we want that to agree with the other case. The adjoint $\partial^*$ has to be constructed to make this identity true, hence this being its definition. Here the adjoint is literally the correct matrix that allows us to get back and forth between pairs of inner products which in our case is pairs of neighboring dimensions.

\section{Combinatorial Hodge Theory}

Given that we know that we can compute the matrix product of the boundary and the coboundary matrix in two ways this can be organized in a way to see what happens in one specific dimension $n$.

\subsection{The Hodge Laplacian}

The first thing to observe about our matrix multiplications of Figure \ref{fig:matrixmultiplication} that the product gets us a square matrix from one vector space to itself so $C_n\rightarrow C_n$. If for example we first took the boundary matrix $\partial_n$ and follow it by the coboundary matrix $\partial_n^T$ we indeed come back to the vector space were we started. By using the inner product to compose, we picked up the metric information along with the topological information in that composition. There are two ways we can go from a given $n$-chain $C_n$: up and down. This is depicted in the following diagram as Figure \ref{fig:hodgelaplacianstructure}:

\begin{figure}
$$ C_{n+1}\stackrel[\partial_{n+1}]{\partial_{n+1}^T}{\longleftrightarrows} C_{n} \stackrel[\partial_{n}]{\partial_{n}^T}{\longleftrightarrows} C_{n-1} $$
\caption{The compositional structure of the Hodge Laplacian. We collect the up connectivity and inner product structure, and the down connectivity and inner product structure into one square matrix on $C_n$.}\label{fig:hodgelaplacianstructure}
\end{figure}

Each matrix product depends on a different boundary matrix. We therefore like to (see Figure \ref{fig:updownlaplacians}) name them separately as up- and down-Laplacians. 

\begin{figure}
$$L_n^{\text{up}} = \partial_{n+1} \partial_{n+1}^T $$
$$L_n^{\text{down}}=  \partial_{n}^T \partial_{n}$$
\caption{The up-Laplacian and the down-Laplacian. In our convention up and down refers to the direction of dimension we are going.}\label{fig:updownlaplacians}
\end{figure}

The full information that can be collected at $C_n$ in each direction can be combined into one square matrix called the {\em Combinatorial Hodge Laplacian} of Figure \ref{fig:combinatorialhodgelaplacian}.

\begin{figure}
$$L_n=L_n^{\text{up}}+L_n^{\text{down}}=\partial_{n+1} \partial_{n+1}^T+\partial_{n}^T \partial_{n}$$
\caption{The Combinatorial Hodge Laplacian}\label{fig:combinatorialhodgelaplacian}
\end{figure}

Given the fundamental lemma of homology, this is all the information we have to collect.

\subsection{Hodge Theorem}

We have constructed the Hodge Laplacian from the same pair of information, two boundary matrices (and their transposes) as we constructed to compute Homology. It should be unsurprising that the following theorem holds (Figure \ref{fig:hodgetheorem}):
\begin{figure}
\[H_n\cong\ker L_n = \ker L_n^{\text{up}} \cap \ker L_n^{\text{down}}\]
\caption{The Hodge Theorem says that the kernel of the Hodge Laplacian captures the homology of the given dimension.}\label{fig:hodgetheorem}
\end{figure}

This theorem is stated in terms of groups, and to convert it to the numbers we love like Betti numbers, we need to take the size of the groups, which for us really just amounts to counting. Say the size of the kernel of the Hodge Laplacian in $0$ dimensions is $2$, we know from Homology, that we have $2$ connected components. Finding the size of the kernel in $1$ dimensions to be $3$ means that we have found 3 loops and so fourth.

The homogeneous solution of the Laplacian operator in vector calculus or partial differential equations is the solution of the zero solution of the Laplacian over acting on a space:

\[L_n x_n=0\]

The solution space has been called {\em harmonic}\marginnote{Solutions of $ker L_n$ are called harmonic.}. This language comes from the homogeneous solution of the usual Hodge Laplacian equation, which is also called harmonic.

\subsection{Hodge Decomposition}

A full discussion of the Hodge Laplacian is found in the Hodge Decomposition. It says that the Hodge Laplacian can be decomposed into three parts: The image of the up-Laplacian, the image of the down-Laplacian and the kernel of the Hodge Laplacian as shown in Figure \ref{fig:HodgeDecomposition}:
\begin{figure}
\[C_n=\rlap{$\underbrace{\phantom{\Ima \partial_{n+1}}}_{\Ima L_n^{\text{up}}}$} \rlap{$\overbrace{ \phantom{\Ima \partial_{n+1}\oplus\ker L_n}}^{\ker L_n^{\text{down}}=\ker \partial_{n}^T}$} \Ima \partial_{n+1}\oplus \underbrace{\ker L_n \oplus\Ima \partial_n^T }_{\ker L_n^{\text{up}}=\ker \partial_{n+1}}\llap{$\overbrace{\phantom{\Ima \partial_{n}^T}}^{\Ima L_n^{\text{down}}}$}\]
\caption{The Hodge Decomposition says that the Hodge Laplacian an be decomposed into three parts.}\label{fig:HodgeDecomposition}
\end{figure}

At first sight, this diagram might be bewildering, but it actually is quite straightforward. The up and down Laplacians contain independent information. We should expect them to be separable. Finally, observe that for this reason, each has to be in the kernel of the other! And this in turn means that the combined kernel is indeed the intersection of the two kernels. That gives us precisely three pieces as described.

If there is no interesting topology, the Hodge decomposition is the same thing as the more familiar Helmholtz decomposition of vector calculus. We will see later examples of this correspondence.

\subsection{The two graph Hodge Laplacians}

Let us study the Hodge Laplacian for graphs. Compare this to our investigation of Homology of graphs. Again, we have just one non-trivial boundary matrix $\partial_1$, but we have two maps involving zeros. Again, we get two ways to place the Hodge Laplacian: $L_0$ and $L_1$ (Figure \ref{fig:twographlaplacians}):

\begin{figure}
\begin{align*}
0\xrightarrow[]{\partial_2} C_1& \xrightarrow[]{\partial_1} C_0 \xrightarrow[]{\partial_0}0 \\
C_1 &\stackrel[\partial_{1}]{\partial_{1}^T}{\longleftrightarrows} C_{0} \stackrel[\partial_{0}]{\partial_{0}^T}{\longleftrightarrows} 0\\
0\stackrel[\partial_{2}]{\partial_{2}^T}{\longleftrightarrows} C_{1} & \stackrel[\partial_{1}]{\partial_{1}^T}{\longleftrightarrows} C_0
\end{align*}
$$L_1:C_1\rightarrow C_1=\Ccancel[red]{L_1^{\text{up}}}+L_1^{\text{down}}=\Ccancel[red]{\partial_{2} \partial_{2}^T}+\partial_{1}^T \partial_{1}=\partial_{1}^T \partial_{1}$$
$$L_0:C_0\rightarrow C_0=L_0^{\text{up}}+\Ccancel[red]{L_0^{\text{down}}}=\partial_{1} \partial_{1}^T+\Ccancel[red]{\partial_{0}^T \partial_{0}}=\partial_{1} \partial_{1}^T$$
\caption{The two Hodge Laplacians of a graph.}\label{fig:twographlaplacians}
\end{figure}
Given the maps involve zeros, some parts of the Hodge Laplacians are zero. We end up with two graph Hodge Laplacians. $L_0$ coincides with a Laplacian that is known in the graph theory literature as the graph Laplacian. $L_1$ is a kind of edge Laplacian. What we see, however, is that a full description of a graph via Hodge Laplacians we do end up with two! This is an example of the benefit of coming to Laplacians via Hodge theory. We are clear about all the information involved and we can make sure we are not missing anything. The two Laplacians are usually different in dimensions. $L_0$ is a square matrix with dimensions given by the number of vertices. $L_1$ is a square matrix with dimensions give by the number of edges. However, both matrices are constructed from the same boundary matrix, so we might expect there to be shared information. We will see this when we discuss spectra.

\subsection{The $L_0$ Graph Laplacian}

One of the two Hodge Graph Laplacians is known as the Graph Laplacian in the graph theory literature. To avoid confusion, we will call this case the $L_0$ graph Laplacian to remind us that there is another one.

Not infrequently the $L_0$ graph Laplacian is defined as follows (Fig. \ref{fig:graphlaplacian}):

\begin{figure}    
\[L_0:=D_G-A_G=\partial_1\partial_1^T\]
\caption{The $L_0$ graph Laplacian defined via Degree and Adjacency Matrices.}\label{fig:graphlaplacian}
\end{figure}

We have not encountered the Degree matrix. In this context, it is defined as the number of edges entering each vertex, or equivalently the number of vertices adjacent to a given vertex. It is a diagonal matrix. A simple example show in Figure \ref{fig:adjvsprod} shows that indeed the matrix derived from the boundary matrix agrees with the one derived from degree and adjacency:
\begin{figure}
\[
L_0=\bordermatrix{
  ~ & e_0 & e_1 & e_2 \cr
  v_0 & \phantom{-}1 & \phantom{-}0 & -1 \cr
  v_1 & -1 & \phantom{-}1 & \phantom{-}0\cr
  v_2 & \phantom{-}0 & -1 & \phantom{-}1\cr
} \cdot \bordermatrix{
  ~ & v_0 & v_1 & v_2 \cr
  e_0 & \phantom{-}1 & -1 & \phantom{-}0 \cr
  e_1 & \phantom{-}0 & \phantom{-}1 & -1\cr
  e_2 & -1 & \phantom{-}0 & \phantom{-}1\cr
}= \bordermatrix{
  ~ & v_0 & v_1 & v_2 \cr
  v_0 & \phantom{-}2 & -1 & -1 \cr
  v_1 & -1 & \phantom{-}2 & -1\cr
  v_2 & -1 & -1 & \phantom{-}2\cr
}
\]

\[
L_0=\bordermatrix{
  ~ & v_0 & v_1 & v_2 \cr
  v_0 & \phantom{-}2 & -1 & -1 \cr
  v_1 & -1 & \phantom{-}2 & -1\cr
  v_2 & -1 & -1 & \phantom{-}2\cr
}=\bordermatrix{
  ~ & v_0 & v_1 & v_2 \cr
  v_0 & 2 & 0 & 0 \cr
  v_1 & 0 & 2 & 0\cr
  v_2 & 0 & 0 & 2\cr
}-\bordermatrix{
  ~ & v_0 & v_1 & v_2 \cr
  v_0 & 0 & 1 & 1 \cr
  v_1 & 1 & 0 & 1\cr
  v_2 & 1 & 1 & 0\cr
}
\]
\caption{$L_0$ Laplacian computed from the boundary matrix $\partial_1$ compared to computed via Degree and Adjacency matrices.}\label{fig:adjvsprod}
\end{figure}

As you see, these two perspectives indeed give identical Laplacians. However, we have so far spent a lot of time deriving interesting information from the boundary matrix. Perhaps a theory could be constructed that would make adjacency/degree matrices as informative, but I am not aware of any such theory. For that reason I think we are currently forced to favor boundary matrix language as being much more informative. This circles back to an earlier suggestion that between adjacency and incidence matrix (which really we now call boundary matrix) we should favor the latter. Dimensional relations are central to understanding homology and this comes directly out of (co)boundary relationships.

\section{Spectra of Laplacians}

The Hodge Laplacians we have constructed are square matrices, but they are also by construction symmetric. Furthermore, given our adjoint construction and the use of the inner product in the computation of the Hodge Laplacian, we also have that the Hodge Laplacian is self-adjoint. 

The eigendecomposition of a matrix transforms a matrix into a diagonalized shape, where the diagonal entries are eigenvalues, and the transforming vectors are eigenvectors. This is a choice of basis of a matrix in which matrix multiplication amounts to scaling in the direction of the eigenvectors. Given that this basis naturally relate to frequencies in oscillatory problems, the eigenvalues are also known as {\em spectra}\marginnote{The set of eigenvalues of a matrix is called its spectrum.}. Put in  equations this looks as follows (Fig. \ref{fig:eigendec}):

\begin{figure}
$$L_n=U_n \Lambda_n U_n^T$$
\[
  \Lambda_n =
  \begin{bmatrix}
    \lambda_{0} & & \\
    & \ddots & \\
    & & \lambda_{\#C_n}
  \end{bmatrix}
\]

\[U_n={u_k}:k\in0,\ldots,\#C_n\]
\caption{The Eigendecomposition of a Laplacian.}\label{fig:eigendec}
\end{figure}

The eigendecomposition of the Hodge Laplacian has some nice properties, because the matrix is symmetric and self-adjoint. A nerdy thing to realize is that self-adjoint matrices are semipositive definite. This means that all eigenvalues are non-negative and real.

\subsection{Why One Graph Laplacian is almost enough}

We have seen earlier that there really are two graph Hodge Laplacians. Yet, graph theory heavily relies on just using one. How much is missed? The following theorem\cite{horak2013spectra} gives us some comfort that graph theorist have been OK. Not much is missed.

\begin{thm}[Horak \& Jost 2011]
    The spectrum $L_0$ and $L_1$ agree, except possibly on the multiplicity of $0$ eigenvalues.
\end{thm}

This of course refers to the spectrum and to the eigenvectors. Those will differ and one will work with a different basis. The next remark hints at how one can inject the missing homological information if one uses only one Hodge Laplacian to study graphs:

\begin{remark}[Corollary of the Hodge Theorem]
The difference in eigenvalue multiplicity is precisely the difference between Betti-0 and Betti-1, that is, the difference between connected components and number of cycles in the graph.
\end{remark}

Despite this, it may still be wisest to recognize that one deals with two Hodge Laplacians for graphs. After all the $L_1$ Hodge Laplacian on graphs naturally captures cycles and will operate on related eigenvectors, while the $L_0$ Hodge Laplacian on graphs naturally captures connected components and will operate on related eigenvector. This keeps the dimensional information clean and intuitive.

\subsection{Simplicial Fourier Transform}

The eigendecomposition of the Hodge Laplacian allows us to define a meaningful notion of a spectral transform theory. In loose association with Fourier Analysis this transform theory was coined the Graph Fourier Transform in the case of graphs and the Simplicial Fourier Transform in the case of simplicial complexes.

Given that the Simplicial Fourier transform contains the graph case as special case and in fact the form of the definition is identical we will just discuss the Simplicial Fourier transform and its inverse in Figure \ref{fig:simplicialfouriertransform}:

\begin{figure}
\begin{align*}
\text{Simplicial Fourier Transform: }\quad& \hat{x}_n=U_n^T x_n\\
\text{Inverse SFT: }\quad& x_n=U_n \hat{x}_n
\end{align*}
\caption{The Simplicial Fourier Transform (SFT) and its inverse.}\label{fig:simplicialfouriertransform}
\end{figure}

Given the eigendecomposition into eigenvectors, the Simplicial Fourier Transform is simply the projection of a $x_n$ vector in an $n$-chain $C_n$ onto the eigenvectors collected in $U_n^T$. The inverse undoes this operation.

\begin{figure}
    \centering
    \includegraphics[width=.95\textwidth]{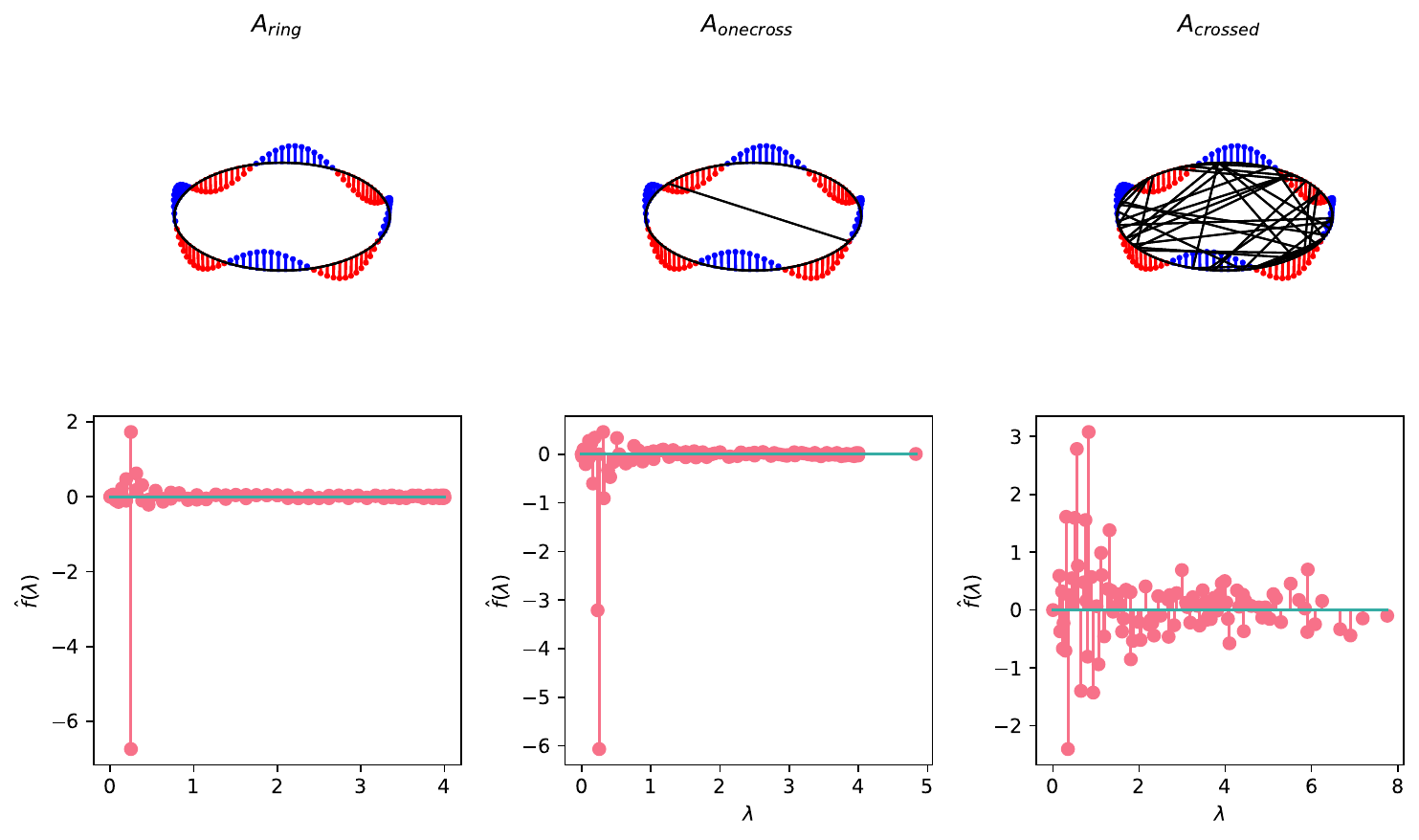}
    \caption{The same signal on a (left) circular topology, (center) circular topology with one cross connection, (right) circular topology with random additional crossconnections.}
    \label{fig:circularcrosslinking}
\end{figure}

The spectrum of the Laplacian picks up both information about the signal and the topology it lives on. Figure \ref{fig:circularcrosslinking} illustrates this. In all cases, the number of vertices are the same and there is a signal on top of them that forms a clean oscillation. In the case of the circular topology, this is picked up clean, but the more cross sections are added the more the topology further influences the spectrum.

\begin{marginfigure}
    \centering
    \includegraphics[width=.95\marginparwidth]{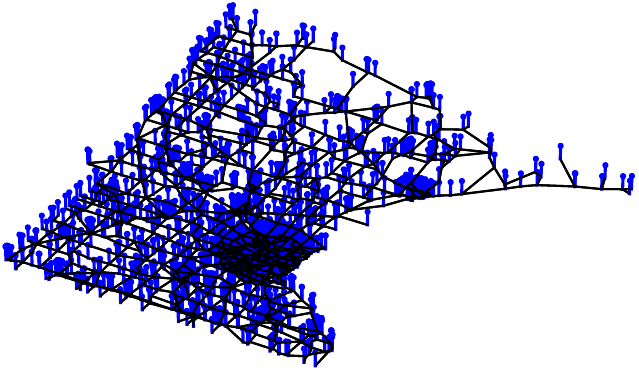}
    \includegraphics[width=.95\marginparwidth]{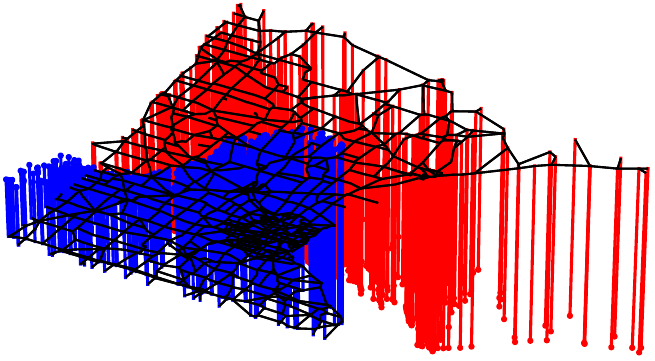}
    \includegraphics[width=.95\marginparwidth]{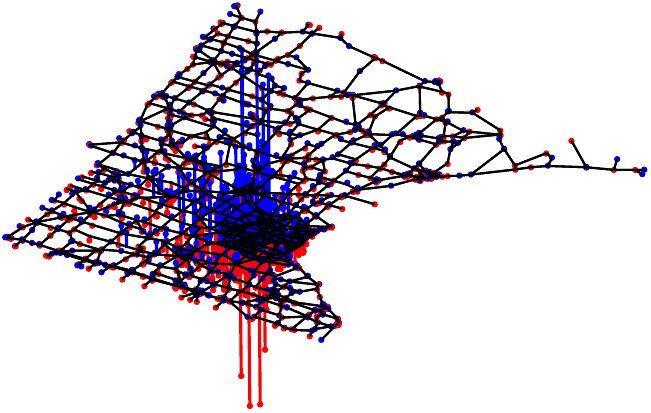}
    \includegraphics[width=.95\marginparwidth]{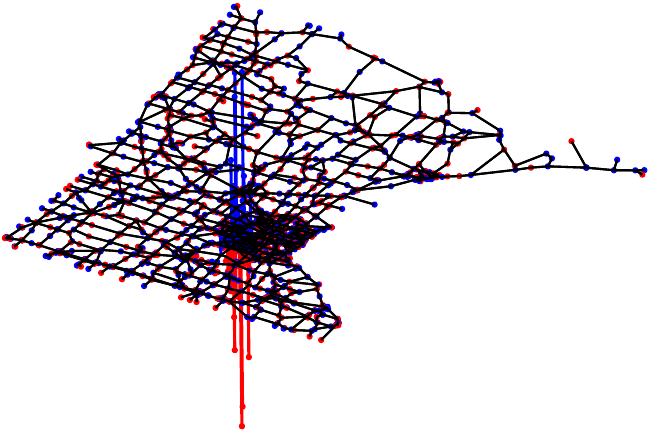}
    \caption{From top to bottom: Zero eigenvalue, first non-zero eigenvalue, eigenvalue at $n/2$, and eigenvalue at $n-1$.}
    \label{fig:minnesota}
\end{marginfigure}

However, in many ways the spectrum of the Laplacian still retains properties that match our more conventional spectral signal representations. To see this, consider the example of the specta of the Minnesota road network dataset\cite{gleich2008}, as included in the graph-signals python library. Figure \ref{fig:minnesota} shows the eigenvectors for the $0$, first non-zero, midpoint and highest eigenvalue. Notice that the "frequency" or undulation of the data goes up as the eigenvalues increase. This means that the increasing eigenvalues do mimic the increasing frequency undulation of the eigenvectors and allowing for interpretations that lean on our standard spectral analysis interpretation of eigenvalues. These examples are mild variations of examples provided as part of the graph-signals python library\cite{bartos2017}.

\section{Signals}

\begin{figure}
     \centering
    \includegraphics[width=.95\textwidth]{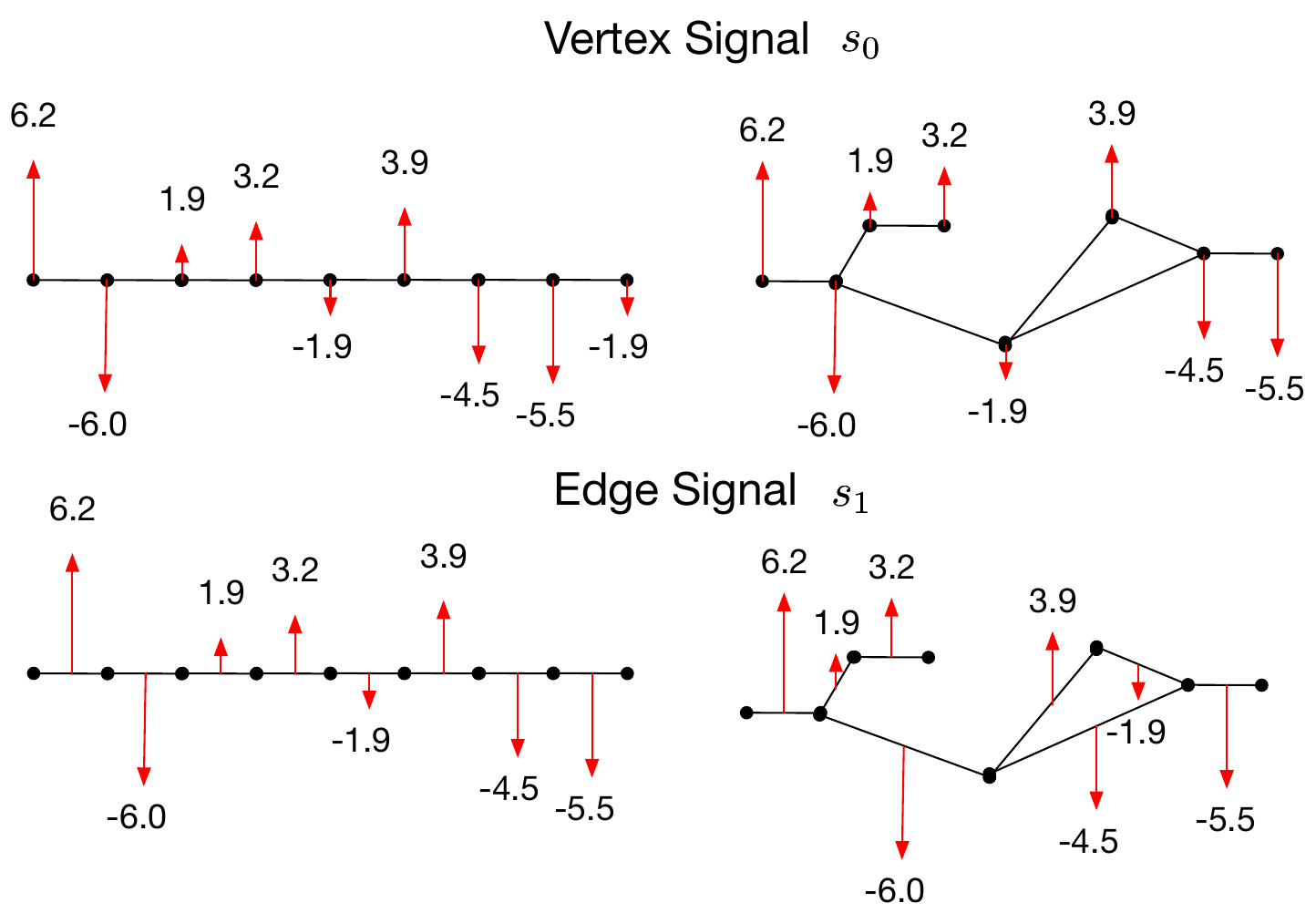}
    \caption{Signals on vertices or edges.}
    \label{fig:vertexedgesignal}
\end{figure}

Signals are usually the thing we care about but the interpretation of signals are domain specific. Generally, we will think of a signal here as some value associated with a simplicial complex. This is justified by considering classical signal notions such as time series on sampling points on the real line as seen in the top left of Figure \ref{fig:vertexedgesignal}. Given that the signal is attached to vertices, we call this type of signal a {\em vertex-signal}. But the same principle can of course be realized over any vertex in a graph or simplicial complex. An example on a graph is seen in the top right of the same figure. We may want to study signals associated with edges, however. This is quite natural and is know under various names, such as edge flow --- or current flow, if we are electrical engineers. It may be meaningful to draw the flow in the direct of the edge, but this is a choice. Here we opt to show {\em edge signals} in the same way we have shown vertex signals, as data attached to the simplex. We see examples over a line graph and a more complex graph in the bottom part of Figure \ref{fig:vertexedgesignal}.

\subsection{Hodge Signal}

The Hodge Decomposition tells us that we have a decomposition into orthogonal subspaces. We can apply this directly to a signal and hence get what we will call a {\em Hodge Signal}.\marginnote{Hodge Signal.}

Hence a signal $s_n$ can be decomposed into three components:
\[s_n = s_n^{\text{irrot}}+s_n^{H}+s_n^{\text{solenoid}}\]
where we lean on terminology from vector calculus to label the components. In terms of boundary and coboundary matrices we can write the same thing as follows giving us a definition to compute each part.
\[s_n=\partial_n^Ts_{n-1}+s^{H}_n+\partial_{n+1} s_{n+1}\]
$$s_n^{\text{irrot}} = \partial^T s_{n-1}$$
$$s_n^{\text{solenoid}} = \partial s_{n+1}$$

\begin{marginfigure}
    \centering
    \includegraphics[width=.95\marginparwidth]{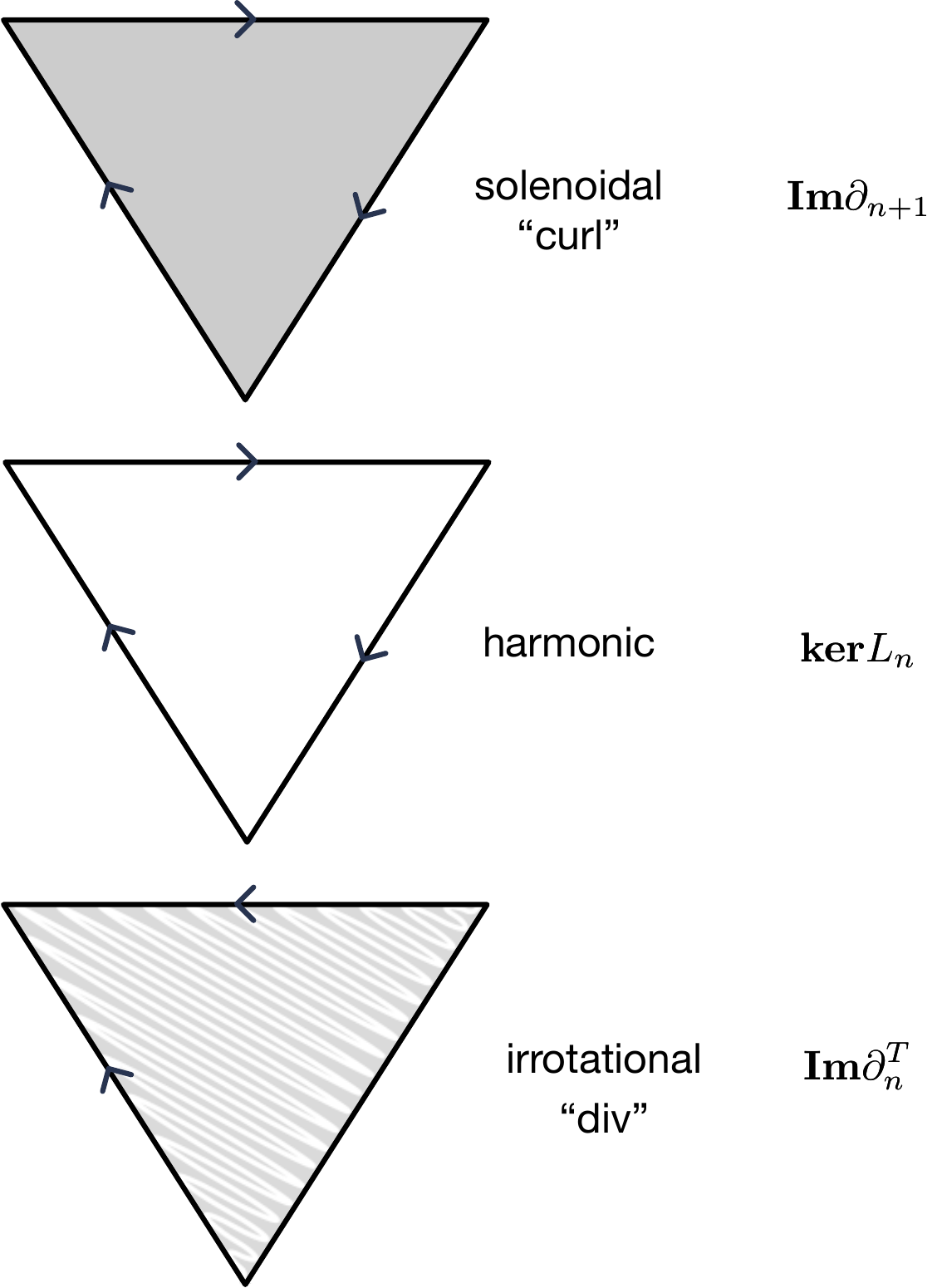}
    \caption{Solenoidal ("curl"), harmonic and irrotational "div" contributions in the combinatorial setting.}
    \label{fig:divharmcurl}
\end{marginfigure}

The Hodge decomposition of a signal on a simplex is depicted in Figure \ref{fig:divharmcurl}.

\section{Simplicial Filtering}

A filter is a linear map that modifies a signal. Take $x_n$ to be the original signal and $y_n$ the filtered signal, and $H$ the linear filter, we write the filter equation:

$$y_n=H x_n$$

Analogous to other filtering settings we want to be able to define a filter as a polynomial over some transform. In digital filters we use z-transforms, and in analog filtering we use continuous Laplace transforms. This motivates the use of our combinatorial Hodge Laplacian to serve as the transform, the discrete Hodge Laplacian transform\marginnote{The Discrete Hodge Laplacian Transform.} in this case and we construct a filter as a polynomial of the Hodge Laplacian.

$$H=\sum_{k=0}^l h_k (L_n)^k=\sum_{k=0}^l h_k (L_n^{\text{up}})^k + \sum_{k=0}^l h_k (L_n^{\text{down}})^k$$

Given that the up and down Laplacian components are independent of each other we can define filter coefficients for each hence allowing us to filter irrotational and solenoidal parts separately.

$$H=\alpha_0 \mathcal{I} + \sum_{j=0}^{l_1} a_j (L_n^{\text{down}})^j + \sum_{k=0}^{l_2} b_k (L_n^{\text{up}})^k$$

$$H=\alpha_0 \mathcal{I} + \sum_{k=0}^l a_k^{\text{irrot}} (L_1^{\text{down}})^k + \sum_{k=0}^l a_k^{\text{solenoid}} (L_1^{\text{up}})^k$$

\subsection{Simplicial Signal Shifts}

In conventional signal processing the powers of the z-transform describe shifts. Shift-like properties can also be recovered for the discrete Hodge Laplacian Transform.

Consider a signal and the same signal shifted $d$ times:
$$n\text{-signal: } s_n\qquad d\text{-shifted }n\text{-signal: } s_n^{d}$$
We can check that the repeated application of the Hodge Laplacian indeed behaves like shifts do for the z-transform, that is repeated applications of the Laplacian acts like a sequence of shifts.
$$L_n^d s_n = \underbrace{L_n L_n \ldots L_n}_d s_n = L_n^1 (L_n^{d-1} s) = L_n^1 s_n^{d-1}$$

\subsection{Filter Properties}

The simplicial filter $H_n$ operating on an n-signal $s_n$ has the following familiar filter properties. It is linear and shift-invariant:

\begin{align*}
\text{Linearity: }H_n (a s_n^1+b s_n^2) &= a H_n s_n^1 + b H_n s_n^2\\
\text{Shift-Invariance: } L_n (H_n s_n) &= H_n (L_n s_n)\\
\end{align*}

Implied by linearity, we also have commutativity:

\[\text{Commutativity: } H_n^{1} H_n^{2} = H_n^{2} H_n^{1}\]

\subsection{Continuous-Combinatorial Correspondence}

We can think of data over our simplicial complex as signals in their own right, but it is interesting to observe that the combinatorial Hodge theory we are using here has a correspondence with the continuous version of Hodge theory.

\subsection{Gradient as Scalar functions on vertices}

A good example to understand this correspondence is as follows. Let us assume a continuous function $f$ that is defined over the vertices as well. If we apply the coboundary matrix to the function, observe that we get {\em finite difference} between the function values at the respective vertices. This of course is the discrete version of the differential of the continuous case.

\[\partial_0^T f = \bordermatrix{
  ~ & v_0 & v_1 & v_2 \cr
  e_0 & \phantom{-}1 & -1 & \phantom{-}0 \cr
  e_1 & \phantom{-}0 & \phantom{-}1 & -1\cr
  e_2 & -1 & \phantom{-}0 & \phantom{-}1\cr
}\cdot \bordermatrix{ 
~ & f \cr
v_0 & f(v_0)\cr
v_1 & f(v_1)\cr
v_2 & f(v_1)\cr
}=\bordermatrix{
~ & "grad" \cr
e_0 & f(v_0)-f(v_1)\cr
e_1 & f(v_1)-f(v_2)\cr
e_2 & f(v_2)-f(v_0)\cr
}\]

{\center $f(\cdot)$ is some function assignment to each vertex in a graph.}

This correspondence goes further and we have continuous version of operators and we have the matrix-based operators we have developed. The following table names an operator and their respective versions in vector calculus and in combinatorial Hodge theory:

\begin{tabular}{c|c|c}
 & continuous & combinatorial \\
 \hline
gradient & $\grad f$ & $\partial_1^T s_0$\\
net node flow & $\Div f$ & $\partial_1 s_1$\\
area rotation & $\curl f$ & $\partial_2^T s_1$\\
rotational flow in an area & $\curl ^* f$ & $\partial_2 s_2$\\
Laplacian & $-\Div \grad$  & $\partial_1\partial_1^T$\\
Helmholtzian & $-\grad \Div + \curl^*\curl$ & $\partial_1^T\partial_1+\partial_2\partial_2^T$\\
\hline
\end{tabular}

\section{Hodge Theory of Sheaves}

It turns out we can fairly easily generalize what we discussed so far to more general conceptions of signals. The way we do this will be via a construction called {\em sheaves}. In our settings sheaves are not complicated, and the generalization will be straightforward.

\subsection{Sheaves for attaching data}

Sheaves provide a way to attach data to a topological space. The mechanism is very general because what we mean here by data is very general. General sheaf theory is much broader than the case we will discuss here. Our sheaves are attached over simplicial complexes, where general sheaves attach to arbitrary topological spaces which can be much more complicated than simplicial complexes. Sheaves as we discuss them here are still very powerful but we avoid plenty of technicalities in this setting.

To get a sense of how sheaves work over a simplicial complex we will now consider a simple example of a {\em line complex} which is the repeated alternation of a $0$-simplex and a $1$-simplex mirroring the pattern of a connected sampled line. 
%
\begin{equation*}
\begin{tikzcd}[scale=1.2,ampersand replacement=\&, column sep=large,/tikz/bullet/.style={circle,fill,inner
sep=2pt,label={[font=\bfseries]#1}},>=latex,execute at end picture={
    \draw [black,line width=1pt,->-/.list={1/2},shorten >=-2.5ex,shorten <=-2.5ex]
    (\tikzcdmatrixname-1-1) node[black,bullet] (a) {}
    -- (\tikzcdmatrixname-1-3) node[black,bullet] (b) {};
    \draw [black,line width=1pt,->-/.list={1/2},shorten >=-2.5ex,shorten <=-2.5ex]     (\tikzcdmatrixname-1-3) node[black,bullet] (b) {}
    -- (\tikzcdmatrixname-1-5) node[black,bullet] (c) {};
  }]
\textcolor{white}{A} \& \textcolor{white}{A} \&\textcolor{white}{A} \&\textcolor{white}{A} \& \textcolor{white}{A}\\[-15pt]
{\mathcal{X}_0} \arrow[r,phantom]{f} \& {\mathcal{X}_1} \&\arrow[l,phantom,swap]{f} \mathcal{X}_0 \arrow[r,phantom]{f} \& {\mathcal{X}_1} \& \arrow[l,phantom,swap]{f} \mathcal{X}_0
\end{tikzcd}
\end{equation*}
%
We can relate simplices via (co)face maps. In this example we pick the coface map $\delta$.\marginnote{Navigations between simplices in a line complex via coface maps.} 
%
\begin{equation*}
\begin{tikzcd}[scale=1.2,ampersand replacement=\&, column sep=large,/tikz/bullet/.style={circle,fill,inner
sep=2pt,label={[font=\bfseries]#1}},>=latex,execute at end picture={
    \draw [black,line width=1pt,->-/.list={1/2},shorten >=-2.5ex,shorten <=-2.5ex]
    (\tikzcdmatrixname-1-1) node[black,bullet] (a) {}
    -- (\tikzcdmatrixname-1-3) node[black,bullet] (b) {};
    \draw [black,line width=1pt,->-/.list={1/2},shorten >=-2.5ex,shorten <=-2.5ex]     (\tikzcdmatrixname-1-3) node[black,bullet] (b) {}
    -- (\tikzcdmatrixname-1-5) node[black,bullet] (c) {};
  }]
\textcolor{white}{A} \& \textcolor{white}{A} \&\textcolor{white}{A} \&\textcolor{white}{A} \& \textcolor{white}{A}\\[-15pt]
{\mathcal{X}_0} \arrow{r}{\delta} \& {\mathcal{X}_1} \&\arrow{l}[swap]{\delta}  \mathcal{X}_0 \arrow{r}{\delta} \& {\mathcal{X}_1} \& \arrow{l}[swap]{\delta}  \mathcal{X}_0
\end{tikzcd}
\end{equation*}
We will denote some data attached to a simplex by $\mathcal{S}$.
\begin{equation*}
\begin{tikzcd}[scale=1.2,ampersand replacement=\&, column sep=large,/tikz/bullet/.style={circle,fill,inner
sep=2pt,label={[font=\bfseries]#1}},>=latex,execute at end picture={
    \draw [black,line width=1pt,->-/.list={1/2},shorten >=-2.5ex,shorten <=-2.5ex]
    (\tikzcdmatrixname-2-1) node[black,bullet] (a) {}
    -- (\tikzcdmatrixname-2-3) node[black,bullet] (b) {};
    \draw [black,line width=1pt,->-/.list={1/2},shorten >=-2.5ex,shorten <=-2.5ex]     (\tikzcdmatrixname-2-3) node[black,bullet] (b) {}
    -- (\tikzcdmatrixname-2-5) node[black,bullet] (c) {};
  }]
\phantom{\cdots} \arrow[r,phantom] \& \phantom{0} \&\arrow[l,phantom] \mathcal{S}_{\phantom{o}} \arrow[r,phantom] \& \phantom{0} \& \arrow[l,phantom] \phantom{\cdots}\\[-15pt]
\textcolor{white}{A} \& \textcolor{white}{A} \&\textcolor{white}{A} \&\textcolor{white}{A} \& \textcolor{white}{A}\\[-15pt]
{\mathcal{X}_0} \arrow{r}{\delta} \& {\mathcal{X}_1} \&\arrow{l}[swap]{\delta}  \mathcal{X}_0 \arrow{r}{\delta} \& {\mathcal{X}_1} \& \arrow{l}[swap]{\delta}  \mathcal{X}_0
\end{tikzcd}
\end{equation*}
The definition of a sheaf requires that we attach data $\mathcal{S}$ to each simplex $\mathcal{X}_\bullet$.\marginnote{Sheaves consist of data $\mathcal{S}$ attached to each simplex.}  
%
\begin{equation*}
\begin{tikzcd}[scale=1.2,ampersand replacement=\&, column sep=large,/tikz/bullet/.style={circle,fill,inner
sep=2pt,label={[font=\bfseries]#1}},>=latex,execute at end picture={
    \draw [black,line width=1pt,->-/.list={1/2},shorten >=-2.5ex,shorten <=-2.5ex]
    (\tikzcdmatrixname-2-1) node[black,bullet] (a) {}
    -- (\tikzcdmatrixname-2-3) node[black,bullet] (b) {};
    \draw [black,line width=1pt,->-/.list={1/2},shorten >=-2.5ex,shorten <=-2.5ex]     (\tikzcdmatrixname-2-3) node[black,bullet] (b) {}
    -- (\tikzcdmatrixname-2-5) node[black,bullet] (c) {};
  }]
\cdots \arrow[r,phantom] \& \mathcal{S} \&\arrow[l,phantom] \mathcal{S}_{\phantom{o}} \arrow[r,phantom] \& \mathcal{S} \& \arrow[l,phantom] \cdots\\[-15pt]
\textcolor{white}{A} \& \textcolor{white}{A} \&\textcolor{white}{A} \&\textcolor{white}{A} \& \textcolor{white}{A}\\[-15pt]
{\mathcal{X}_0} \arrow{r}{\delta} \& {\mathcal{X}_1} \&\arrow{l}[swap]{\delta}  \mathcal{X}_0 \arrow{r}{\delta} \& {\mathcal{X}_1} \& \arrow{l}[swap]{\delta}  \mathcal{X}_0
\end{tikzcd}
\end{equation*}
A sheaf construction further requires that whenever there is a map between simplices, we have to provide a map between the attached data.\marginnote{Sheaves also require that for each map between simplices we provide a map between sheaf data.}  
%
\begin{equation*}
\begin{tikzcd}[scale=1.2,ampersand replacement=\&, column sep=large,/tikz/bullet/.style={circle,fill,inner
sep=2pt,label={[font=\bfseries]#1}},>=latex,execute at end picture={
    \draw [black,line width=1pt,->-/.list={1/2},shorten >=-2.5ex,shorten <=-2.5ex]
    (\tikzcdmatrixname-2-1) node[black,bullet] (a) {}
    -- (\tikzcdmatrixname-2-3) node[black,bullet] (b) {};
    \draw [black,line width=1pt,->-/.list={1/2},shorten >=-2.5ex,shorten <=-2.5ex]     (\tikzcdmatrixname-2-3) node[black,bullet] (b) {}
    -- (\tikzcdmatrixname-2-5) node[black,bullet] (c) {};
  }]
\cdots \arrow[r] \& \mathcal{S} \&\arrow[l] \mathcal{S}_{\phantom{o}} \arrow[r] \& \mathcal{S} \& \arrow[l] \cdots\\[-15pt]
\textcolor{white}{A} \& \textcolor{white}{A} \&\textcolor{white}{A} \&\textcolor{white}{A} \& \textcolor{white}{A}\\[-15pt]
{\mathcal{X}_0} \arrow{r}{\delta} \& {\mathcal{X}_1} \&\arrow{l}[swap]{\delta}  \mathcal{X}_0 \arrow{r}{\delta} \& {\mathcal{X}_1} \& \arrow{l}[swap]{\delta}  \mathcal{X}_0
\end{tikzcd}
\end{equation*}
Notice that typically we have more than one map pointing to the same data $\mathcal{S}$. This leads to a final requirement for sheaves. Informally we have to avoid that there is a conflict between these two maps. They have to in some suitable sense agree what $\mathcal{S}$ is. This can be thought of in two ways. The first is via composition. One can require that the two maps can be {\em composed}. Another way to think about this is to require that local data has to be {\em consistent}. With these three rules we have a full definition of a sheaf over a simplicial complex. In examples we will see how that is realized in practice soon.\marginnote{Sheaves maps are required to allow composition. Alternatively we can think of local sheaf data being required to be consistent.}  
%
\begin{equation*}
\begin{tikzcd}[scale=1.2,ampersand replacement=\&, column sep=large,/tikz/bullet/.style={circle,fill,inner
sep=2pt,label={[font=\bfseries]#1}},>=latex,execute at end picture={
    \draw [black,line width=1pt,->-/.list={1/2},shorten >=-2.5ex,shorten <=-2.5ex]
    (\tikzcdmatrixname-2-1) node[black,bullet] (a) {}
    -- (\tikzcdmatrixname-2-3) node[black,bullet] (b) {};
    \draw [black,line width=1pt,->-/.list={1/2},shorten >=-2.5ex,shorten <=-2.5ex]     (\tikzcdmatrixname-2-3) node[black,bullet] (b) {}
    -- (\tikzcdmatrixname-2-5) node[black,bullet] (c) {};
    \node [rounded corners,draw,dashed,
    inner xsep=-1pt, blue,line width=1.5pt,fit={(\tikzcdmatrixname-1-1) (\tikzcdmatrixname-1-2)}]{};
    \node [rounded corners,draw,dashed,
    inner xsep=-1pt, green,line width=1.5pt,fit={(\tikzcdmatrixname-1-2) (\tikzcdmatrixname-1-3)}]{};
  }]
\cdots \arrow[r] \& \mathcal{S} \&\arrow[l] \mathcal{S}_{\phantom{o}} \arrow[r] \& \mathcal{S} \& \arrow[l] \cdots\\[-15pt]
\textcolor{white}{A} \& \textcolor{white}{A} \&\textcolor{white}{A} \&\textcolor{white}{A} \& \textcolor{white}{A}\\[-15pt]
{\mathcal{X}_0} \arrow{r}{\delta} \& {\mathcal{X}_1} \&\arrow{l}[swap]{\delta}  \mathcal{X}_0 \arrow{r}{\delta} \& {\mathcal{X}_1} \& \arrow{l}[swap]{\delta}  \mathcal{X}_0
\end{tikzcd}
\end{equation*}
\section{Sheaf Cohomology}

It turns out that if our sheaf data are vectors from finite vector spaces and our sheaf maps are linear maps, one can define a cohomology of sheaves. This might be believable now simply because we would be dealing with linear maps of which we can look at image and kernel and we have maps that relate to going in the same direction as the coface of a simplex. But more importantly what is an intuition what sheaf cohomology captures? After all we no longer just build up matrices to capture the simplicial connectivity and how it relates to forming of voids.

\subsection{Consistency in a Shift Register}

Sheaf cohomology is nicely illustrated with a simple example. Here we consider a line complex with a shift map that shifts a vector to the left.

\[
[1,2,\colorbox{green}{3}]\xrightarrow[]{\begin{bsmallmatrix}0&1&0\\0&0&1\end{bsmallmatrix}}[2,3]\xleftarrow[]{\begin{bsmallmatrix}1&0&0\\0&1&0\end{bsmallmatrix}}[2,3,\colorbox{green}{4}]\xrightarrow[]{\begin{bsmallmatrix}0&1&0\\0&0&1\end{bsmallmatrix}}[3,4]\xleftarrow[]{\begin{bsmallmatrix}1&0&0\\0&1&0\end{bsmallmatrix}}[3,4,\colorbox{green}{5}]
\]

Notice that the way we direct the shifts follow the coface map on the underlying line simplex. Two such maps point onto the same data. That data has to agree to be consistent. A linear algebraic way to describe that the data from two sources is identical is to take its difference. If that difference is zero then indeed they do agree. This observation we encode in the following matrix equation. The red block is the linear map coming from the left and the blue block is the linear map coming from the right. Observe that here we chose the blue blocks to have the negative sign. If we now multiply these blocks with their assignment, which is the vector multiplied on the right, we should always get zero if this assignment is indeed consistent.

\[
\begin{bNiceMatrix}
  
\Block[fill=red!15,rounded-corners]{2-3}{}0 & 1 & 0 & \Block[fill=blue!15,rounded-corners]{2-3}{}-1 &  \phantom{-}0 & \phantom{-}0 & \phantom{-}0 & \phantom{-}0 & 0  \\
0 & 0 & 1 & \phantom{-}0 & -1 & \phantom{-}0 & \phantom{-}0 & \phantom{-}0 & 0  \\
0 & 0 & 0 & \Block[fill=red!15,rounded-corners]{2-3}{}\phantom{-}0 & \phantom{-}1 & 0 & \Block[fill=blue!15,rounded-corners]{2-3}{}-1 &  \phantom{-}0 & 0 \\
 0 & 0 & 0 & \phantom{-}0 & \phantom{-}0 & \phantom{-}1 &  \phantom{-}0 & -1 & 0  \\
\end{bNiceMatrix}
\cdot\begin{bNiceMatrix} \Block[fill=green!5,rounded-corners]{3-1}{}1 \\ 2 \\ \Block[fill=green!50,rounded-corners]{1-1}{}3 \\ \Block[fill=green!5,rounded-corners]{3-1}{}2 \\ 3 \\ \Block[fill=green!50,rounded-corners]{1-1}{}4 \\ \Block[fill=green!5,rounded-corners]{3-1}{}3 \\ 4 \\ \Block[fill=green!50,rounded-corners]{1-1}{}5 \end{bNiceMatrix}=\begin{bmatrix} 0 \\ 0\\ 0 \\ 0 \end{bmatrix}
\]

It is worthwhile checking this in a few places. For example take the number $2$ in the second position from the top in the vector. In the top row of our matrix its multiplied by 1 and the second $2$ in the vector is multiplied by $-1$ so indeed we get zero. This works out for most elements in the vector. However, some entries in the vector ($1$ and $5$) always multiply by zero! These are assignments that are arbitrary and hence are always consistent. We could pick any other assignments and would still get zero. 

\subsection{Coboundary structure of the Consistency construction}

If we remove the details of the sheaf maps represented as vectors and just keep the sign, we get this matrix structure:

\[
\underbrace{\begin{bNiceMatrix}
\Block[fill=red!15,rounded-corners]{2-3}{1}\phantom{0} & \phantom{1} & \phantom{0} & \Block[fill=blue!15,rounded-corners]{2-3}{-1}\phantom{-1} &  \phantom{-0} & \phantom{-0} & \Block{2-3}{0}\phantom{-0} & \phantom{-0} & \phantom{0}  \\
\phantom{0} & \phantom{0} & \phantom{1} & \phantom{-0} & \phantom{-1} & \phantom{-0} & \phantom{-0} & \phantom{-0} & \phantom{0}  \\
\Block{2-3}{0}\phantom{0} & \phantom{0} & \phantom{0} & \Block[fill=red!15,rounded-corners]{2-3}{1}\phantom{-0} & \phantom{-1} & \phantom{0} & \Block[fill=blue!15,rounded-corners]{2-3}{-1}\phantom{-1} &  \phantom{-0} & \phantom{0} \\
 \phantom{0} & \phantom{0} & \phantom{0} & \phantom{-0} & \phantom{-0} & \phantom{-1} &  \phantom{-0} & \phantom{-1} & \phantom{0}  \\
\end{bNiceMatrix}}_{\delta^\mathcal{S}_0}
\cdot\begin{bNiceMatrix} \Block[fill=green!45,rounded-corners]{3-1}{}\phantom{1} \\ \phantom{2} \\ \phantom{3} \\ \Block[fill=green!15,rounded-corners]{3-1}{}\phantom{2} \\ \phantom{3} \\ \phantom{4} \\ \Block[fill=green!30,rounded-corners]{3-1}{}\phantom{3} \\ \phantom{4} \\ \phantom{5} \end{bNiceMatrix}=0
\]

This really should be familiar. It is the same structure that a coboundary matrix has in cohomology with the sign capturing orientation. This is why we can call this large matrix $\delta^\mathcal{S}_0$ for the zero-th sheaf coboundary matrix. This works in all dimensions, so we get sheaf coboundary maps for dimension $n$ written as $\delta^\mathcal{S}_n$. While in cohomology we like to use superscripts to indicate cohomology, here we chose to put the sheaf in superscript. This is merely notational convention and one might just flip them if that feels more comfortable.

\marginnote{Sheaf Cohomology}

\[H^n(\mathcal{X},\mathcal{S})=\ker \delta^\mathcal{S}_n/\Ima \delta^\mathcal{S}_{n-1}\]

Sheaf cohomology is a richer object than standard cohomology because we now have both the topological structure of the simplicial complex and the data on top of it jointly forming the linear maps from which we compute cohomology. It is however a good intuition that sheaf cohomology captures data consistency in all dimensions, and this can be most sensibly understood in low dimensions. The zero-th sheaf cohomology $H^0(\mathcal{X},\mathcal{S})$ is the space of global assignments. The first sheaf cohomology $H^1(\mathcal{X},\mathcal{S})$ captures closed oriented data loops such as resonances or feedback. Higher dimensional sheaf cohomology is harder to think about, but we can guess that it is data organized around $n$-voids in a consistent way such that we no longer look at data already accounted for in $n-1$ dimensions.

\subsection{Sheaf Laplacians}

Perhaps it is now no surprise that we can now define a {\em Sheaf Laplacian} in the exact same way that we have already defined the combinatorial Hodge Laplacian over a simplicial complex. After all, all we did is take transposes (adjoints) and have an inner product handy. That will again allow us to matrix multiply and have all the machinery ready to define a Laplacian. Given that this Laplacian is based on the sheaf coboundary matrices rather than the coboundary matices, it can be thought of as a direct generalization. We write the sheaf Laplacian as follows:

\[
L_n(\mathcal{X},\mathcal{S},\langle,\rangle)=\delta^\mathcal{S}_n(\delta_n^\mathcal{S})^*+(\delta_n^\mathcal{S})^*\delta_n^\mathcal{S}
\]

where $\mathcal{X}$ is still the simplicial complex, $\mathcal{S}$ is the sheaf and $<,>$ are the chosen inner products. All these are still just linear maps and we have kernels and images, and in fact we have the Sheaf Hodge theorem:

\[
\ker L_n(\mathcal{X},\mathcal{S},\langle,\rangle)\cong H^n(\mathcal{X},\mathcal{S})
\]
If we take the constant sheaf we recover the simplicial Laplacian!

\section{Parting Observations}

If you paid close attention, we really only used linear algebra to do any computation. To compute (co)homology we construct matrices from combinatorial objects and then compute their rank (and rank deficiency). For Hodge Laplacians we add an inner product and compute matrix multiplications with the matrix transpose. Computationally sheaf cohomology and sheaf Laplacians turn out to actually not use anything else either. They just use matrices in the place where the ordinary Hodge Laplacian used coefficients. So, in summary it is probably fair to say that all we use is fairly elementary linear algebra\marginnote{We just needed linear algebra to do all this!}! We have seen a bit of group theory to write down homology, but it really was just an alternative exposition and not required for computation. One can safely skip these and not lose the ability to understand and compute anything up to sheaf Laplacians. This is good news, because this beautiful theory is actually fairly easy to understand and easy to apply. Topics that are essentially linear algebraic in nature such as linear time-invariant filter theory or linear control theory can immediately interface with the topic. We have seen the discrete side of a correspondence of continuous and discrete theories. This is important to fields that use numerical approximation of continuous geometetries via finite difference and finite element methods including computer graphics. The interested reader should consult a good book on Riemannian geometry get the continuous side of this story in more detail.

\section{Epilogue --- More to know that we did not cover}

The fields of topological signal processing and topological data analysis are much richer than could be covered here. For example topics such as {\em Topological Features in Machine Learning} have all been left out completely. Other topics such as quotient topologies and their relationship to periodic phenomena have been discussed in previously published lecture notes \cite{essl2022topology}, as has persistent homology, time-series embedding and a few other ideas, such as using sheafs not just for linear data and maps but for non-linear ones as well such as feedback frequency modulation. 

\subsection{Vistas for future research}

Despite its substantial history topological data analysis and topological signal processing have left many open research avenues ripe for exploration. In particular in the realm of audio signal processing there remains much to be done. Here are but two of many possible topics of interest: {\em Topological Harmonic Analysis - Build bridge between Persistence and Fourier}, {\em Sheaves over higher order topologies}.

\subsection{Further Reading}
To learn more details and cover gaps in the material discussed here, the following are good yet fairly accessible long form treatments:
\begin{itemize}
\item Ortega, Intro to Graph Signal Processing, 2022.
\item Battison and Petri (eds), Higher-Order Systems, 2022.
\item Robinson, Topological Signal Processing, 2014.
\item Barbarossa \& Sardellitti. "Topological signal processing over simplicial complexes." IEEE TSP (68), 2020.
\end{itemize}

\subsection{Software Recommendations}
The landscape of software implementation of topological algorithms is already fairly vast and rapidly expanding. This is a very small selection curated for either ease of use, relevance to signal processing, or performance, and connected to the material presented in these notes is listed below:
\begin{itemize}
\item graph-signals (Pyhton) - Graph Signal Processing
\item pygsp (Python) - Graph Signal Processing
\item GraSP (Mathlab) - Graph Signal Processing
\item pyDEC (Python) - Discrete Exterior Calculus (including Hodge Theory)
\item pysheaf (Python) - Sheaves over Simplicial Complexes
luding Topological Signal Processing
\end{itemize}

\section{Acknowledgements}
I owe Stefania Serafin, chair of DAFx-2023 my gratitude for offering the time to present this material during as a tutorial at the conference.  These notes have benefited from feedback of numerous tutorial attendees. Last but not least the material presented is part of a larger book project with the generous support of a Simon Guggenheim Foundation fellowship.
Figure \ref{fig:simplicialcomplex} is a public domain figure from wikipedia.

\bibliography{main.bib} 
\bibliographystyle{plainnat}

\end{document}